\documentclass[11pt]{article}
 \usepackage{mathtools}
 \usepackage{adjustbox}
 
 \usepackage[iso-8859-7]{inputenc}
 \usepackage{epsfig}
 \usepackage{array}
 \newcolumntype{P}[1]{>{\centering\arraybackslash}p{#1}}
 \usepackage{graphicx}
 \usepackage{epstopdf}
\usepackage{amsfonts}
 \usepackage{amssymb}
 \usepackage{amsthm}
 \usepackage{amsmath}
 \usepackage{multirow}
 \usepackage{color}

 \newcommand{\nn}{{\nonumber}}
 \newcommand{\eps}{{\epsilon}}
 \newcommand{\beq}{\begin{equation}}
 	\newcommand{\eeq}{\end{equation}}
 \newcommand{\bea}{\begin{eqnarray}}
 	\newcommand{\eea}{\end{eqnarray}}
 
 \newcommand{\gsim}{\lower.7ex\hbox{$\;\stackrel{\textstyle>}{\sim}\;$}}
 \newcommand{\lsim}{\lower.7ex\hbox{$\;\stackrel{\textstyle<}{\sim}\;$}}
 \newcommand{\be}{\begin{equation}}
 	\newcommand{\ee}{\end{equation}}
 \newcommand{\ba}{\begin{eqnarray}}
 	\newcommand{\ea}{\end{eqnarray}}

 \usepackage{cite}
 \usepackage{float}
 \usepackage{amsmath} 
 \usepackage{amssymb}   
 \usepackage{amsthm} 
 \usepackage{geometry}
 \usepackage{graphicx} 
 \usepackage{multirow} 
 \title{Summary of stabilization}

 \usepackage{scalerel}
\usepackage{tikz}
\usetikzlibrary{svg.path}

\definecolor{orcidlogocol}{HTML}{A6CE39}
\tikzset{
  orcidlogo/.pic={
    \fill[orcidlogocol] svg{M256,128c0,70.7-57.3,128-128,128C57.3,256,0,198.7,0,128C0,57.3,57.3,0,128,0C198.7,0,256,57.3,256,128z};
    \fill[white] svg{M86.3,186.2H70.9V79.1h15.4v48.4V186.2z}
                 svg{M108.9,79.1h41.6c39.6,0,57,28.3,57,53.6c0,27.5-21.5,53.6-56.8,53.6h-41.8V79.1z M124.3,172.4h24.5c34.9,0,42.9-26.5,42.9-39.7c0-21.5-13.7-39.7-43.7-39.7h-23.7V172.4z}
                 svg{M88.7,56.8c0,5.5-4.5,10.1-10.1,10.1c-5.6,0-10.1-4.6-10.1-10.1c0-5.6,4.5-10.1,10.1-10.1C84.2,46.7,88.7,51.3,88.7,56.8z};
  }
}

\newcommand\orcidicon[1]{\href{https://orcid.org/#1}{\mbox{\scalerel*{
\begin{tikzpicture}[yscale=-1,transform shape]
\pic{orcidlogo};
\end{tikzpicture}
}{|}}}}

 \usepackage{hyperref}

\begin{document} 
 	\thispagestyle{empty}
 	\begin{titlepage}
 		\vspace*{0.7cm}
 		\begin{center}
 			{\Large {\bf  Remarks on the effects of quantum corrections on moduli stabilization and de Sitter vacua in type IIB string theory }}
 			\\[12mm]
 			Vasileios Basiouris~\orcidicon{0000-0002-9266-7851}~\footnote{E-mail: \texttt{v.basiouris@uoi.gr}}, 
 			George K. Leontaris~\orcidicon{0000-0002-0653-5271}~\footnote{E-mail: \texttt{leonta@uoi.gr}} 
 		\end{center}
 		\vspace*{0.50cm}
 		\centerline{\it
 			Physics Department, University of Ioannina}
 		\centerline{\it 45110, Ioannina, Greece}
 		\vspace*{1.20cm}
 		
 		\begin{abstract}
 		The r\^ole of string loop corrections on the existence of de Sitter vacua  and the moduli stabilization problem is examined in type IIB effective theories. The fundamental building blocks are a minimum of three intersecting D7 brane stacks, three K\"ahler moduli, and a novel Einstein-Hilbert term associated with higher derivative terms of  the 10-dimensional effective action. It was shown in previous works that  loop corrections appear which induce novel  logarithmic volume-dependent terms in the effective potential. When D-term contributions are considered, all K\"ahler moduli are stabilized and de Sitter vacua are achieved. In the present work, a comprehensive study of multiple non-perturbative terms in the superpotential is undertaken.  The combined effects of the logarithmic loop corrections and two non-perturbative superpotential K\"ahler moduli dependent terms have been investigated.  It is shown that a variety of fluxes exist for large as well as moderate volume compactifications which   define a de Sitter space and stabilize the moduli fields. For large volumes, a generic simple form of the potential is achieved. The so obtained effective potential appears to be promising for cosmological applications.
 
 		\end{abstract}
 		
 	\end{titlepage}

 \newpage 
 	 	\section{Introduction}
 	 	There  has been a lot of recent activity regarding the implications of quantum corrections to the moduli stabilization problem and the existence of de Sitter (dS) vacua in effective string theory models. These investigations are significant for providing an answer of whether the otherwise presumably consistent effective theories have an ultraviolet completion, and as such, they could be accommodated in the string landscape\footnote{There is a vast literature on this issue.  For the recent debate on Swampland  conjectures see for example
 	 	~ \cite{Vafa:2005ui,Obied:2018sgi,Agrawal:2018own}  and related reviews~\cite{Palti:2019pca,vanBeest:2021lhn,Danielsson:2018ztv}. 
      Also, an incomplete list of 10d supergravity dS  solutions and other related issues is~\cite{Andriot:2021rdy,Andriot:2020vlg,Basile:2021krr,Basile:2021krk,Roupec:2018mbn,Junghans:2018gdb,Blumenhagen:2020doa}}. These explorations  are also  of great importance since they are related to  the accelerated expansion of the universe and the scenario of cosmological inflation.

 	An appropriate framework to investigate the effects of quantum corrections is type IIB string theory and, more generally, F-theory  defined on an elliptically fibered Calabi-Yau (CY) fourfold. The  fundamental constituents of the effective field theory (EFT) emerging after  compactification of the ten-dimensional theory to four dimensions on a CY manifold, are the  superpotential ${\cal W}$ and the  K\"ahler potential ${\cal K}$ which is of no-scale type. The basic ingredients are the various types of moduli fields associated with the various deformations of the compactification, D-brane stacks with  magnetic fluxes and topological parameters such as the Euler characteristic.  At the classical level of the resulting EFT a number of pertinent issues arise. The tree-level superpotential is a function of the various types of those fields, including the axion-dilaton and the complex structure  moduli, however, it does not depend on the K\"ahler ones. Supersymmetric	conditions imposed on ${\cal W}$ in principle can fix the values of all but the K\"ahler moduli.	On the other hand,	the scalar potential vanishes identically due to the no scale structure of the K\"ahler potential and as a consequence, the K\"ahler	moduli remain undetermined. Moreover, a
 	vanishing scalar potential cannot describe the accelerated  expansion of the universe which requires a positive cosmological constant, that is, a dS minimum. 
 	 	
 	   The  way to confront these issues is to incorporate quantum corrections in ${\cal W}$ and ${\cal K}$ functions.  The superpotential  ${\cal W}$ receives  non-perturbative 
 	   corrections appearing through exponential terms which  depend on the  K\"ahler (volume) moduli~(see for example~\cite{Kachru:2003aw} and~\cite{Cicoli:2007xp})\footnote{For recent developments on related work see \cite{Polchinski:2015bea,Hamada:2019ack,Carta:2019rhx,Blumenhagen:2019qcg,Demirtas:2019sip,Alvarez-Garcia:2020pxd,Broeckel:2021uty,Honma:2021klo}}.   In the context of type IIB string theory the  K\"ahler potential receives perturbative corrections  where the leading order is  ${\alpha'}^3$~\cite{Becker:2002nn}. 
 	    Recently,  the effects of  gravitational terms beyond the standard Einstein-Hilbert term have been considered in the
 	     context of a geometric configuration involving three $ D7$ brane stacks intersecting each other.  The terms next to leading order
 	   in the low-energy expansion of type II superstring are fourth order (${\cal R}^4$) in the Riemann curvature ${\cal R}$, which do not receive any perturbative
 	   corrections beyond one loop~\cite{Grisaru:1986px,Antoniadis:1997eg,Antoniadis:2003sw,Green:1997di,Kiritsis:1997em}.
 	    Dimensional reduction of the  ten-dimensional string effective action in  four dimensions
 	    induces an additional  Einstein-Hilbert term appearing in the bulk multiplied by  a factor proportional to the Euler characteristic.  
      The amplitude induced by graviton scattering involving two massless gravitons	and a Kaluza-Klein (KK) excitation propagating towards a $D7$ brane stack yields 
 	   logarithmic contributions breaking  the no scale invariance of the	K\"ahler potential~\cite{Antoniadis:2018hqy,Antoniadis:2019rkh}.  This induces an F-term  anti de Sitter scalar potential whereas  $U(1)$ symmetries associated with the $D7$ brane stacks 	provide the necessary uplift to a dS minimum 
 	   through  positive D-term  contributions to the scalar potential.
 	Both, logarithmic corrections and D-terms are sufficient to stabilize the K\"ahler
 	moduli and  support a positive cosmological constant.  Also, the implications of this construction  on the cosmological inflation have been studied. It was found that hybrid inflation is successfully implemented with  the internal volume modulus acting as  the inflaton~\cite{Antoniadis:2020stf},
 	and the r\^ole of the waterfall  field is played by open string excitations associated with the $D7$ brane stacks~\cite{Antoniadis:2021lhi}.

 	Within the above context, in a previous work~\cite{Basiouris:2020jgp} a first step towards combining the effects of the
 logarithmic (perturbative) and the non-perturbative corrections  has been put forward.    The analysis  was performed within a
 framework of three K\"ahler moduli $\rho_i$ and three intersecting $D7$ brane stacks and,
 on first  approach, only one K\"ahler modulus $\rho_1$  has been  considered to contribute 
 in the flux induced superpotential ${\cal W}_0$ through an exponential term of the form $Ae^{-a \rho_1}$. 
 A circumstantial study of the scalar potential revealed  regions of the fluxes and other  parameters~\footnote{While several  properties  of the effective model emerging from  String theory   are already fixed, yet, there are  free  parameters such as the Euler characteristic related to the compactification manifold 
 	and the various magnetic fluxes  on $ D7$ branes  which  define 
 the final shape of	Quantum Corrections. }  supporting dS minima in agreement with 
 the present day cosmological observations.

  Usually, however, in realistic compactification scenarios  more complicated 
  situations arise where several K\"ahler moduli induce  non-perturbative terms in the superpotential.
  	In the present work, a useful extension of the previous analysis with two such terms (involving two distinct  K\"ahler moduli)  is performed,  which captures many features of previous models. 
For example, an analogous situation with two exponential terms arises where such type of 
exponential terms in the superpotential   are introduced when  	E3-instantons wrap  two $dP_5$ cycles\cite{Cicoli:2021dhg}.  
Also, in the limiting case where the two exponentials depend on the same K\"ahler modulus, the present construction reduces to the particular  
racetrack paradigm studied in the past~\cite{Blanco-Pillado:2004aap}.

Taking into account the above remarks,  the objective  of the subsequent analysis is the investigation of the combined effects of perturbative and
non-perturbative contributions to the scalar potential of the effective theory.
Regarding the perturbative loop corrections, of particular interest
in this  work are those related to novel graviton kinetic terms in the
bulk which receive logarithmic corrections due to the emission of closed strings propagating in a two dimensional space towards $D7$ probes. In the K\"ahler potential (denoted with $\mathcal{K}$), these corrections appear as a shift to the internal volume ${\cal V}$,  breaking its no scale structure whereas this logarithmic
dependence of the K\"ahler  (volume) moduli is conveyed to the  scalar potential of the effective theory through $\mathcal{K}$.  Hence, these corrections  imply a structure of the potential -analogous to that realizing the Coleman-Weinberg mechanism~\cite{Coleman:1973jx}-  which depends 
on the K\"ahler moduli and only a few  parameters  such as the magnetic    fluxes and topological  invariants of the compactification manifold.

In the following section, (section 2), the basic features of the model together
with a  short review of the previous work~\cite{Antoniadis:2018hqy,Antoniadis:2019rkh,Basiouris:2020jgp} are presented.
In section 3, the scalar potential is computed taking 
into account  the aforementioned  perturbative logarithmic contributions and the non-perturbative corrections. 
 In section 4, D-term contributions are included. A simple form of the total
potential is presented in the large volume regime which shows that all 
moduli directions are stabilized along a metastable dS minimum. 
Section 5 deals with the search of dS solutions for moderate values 
of the K\"ahler moduli.  A summary of the work and conclusions are described in section 7 and some computational details are found in the Appendix.

 	\section{A short review and extension of previous work  }

 	In~\cite{Basiouris:2020jgp}  a  model  consisting of a geometric configuration of three $D7$ branes and  three 	
 	K\"ahler moduli  based on the construction proposed in~\cite{Antoniadis:2018hqy} in the framework of type IIB string theory,
 	 has been studied beyond the tree-level approximation, by including  logarithmic perturbative~\cite{Antoniadis:2019rkh} as well as non-perturbative corrections.   
 	 Despite the complicated  structure	of these contributions, it was shown that,  in the large volume regime, the scalar potential 
 	 of the emerging effective field theory receives a simplified form which illustrates all the essential features of the model.
 	 Within this context, the properties of the effective potential regarding the K\"ahler moduli stabilization and the search for de Sitter vacua
 	 have been studied and the main findings are recapitulated here. Starting with notations and conventions, 	the K\"ahler moduli  
 	 are denoted with
 	\be 
 	\rho_k= b_k+i \tau_k, \;k=1,2,3~, \label{3rhos}
 	\ee  
 	where  $b_k $  are related to the RR $C_4$-form potential and $\tau_k$ are four-cycle volumes. In terms of the latter,
 	in the present work, the internal volume is written as follows
 	\be 
 	{\cal V} = \sqrt{\tau_1 \tau_2 \tau_3}~\cdot\label{svol}
 	\ee
 	The flux induced  superpotential at the classical level depends on  the complex structure moduli $ z_a$, and is given by the Gukov-Vafa-Witten
 	formula~\cite{Gukov:1999ya}  
 	\be 
 	 {\cal W}_0=  \int\, G_3\wedge { \Omega}( z_a)~.\label{GVW} 
 	 \ee  
 	 The symbol $G_3$  represents  the combination $F_3-S H_3$ of the field 
 	strengths $F_3=dC_2,H_3=dB_2$  and  the axion-dilaton modulus $S=C_0+ie^{-\phi}$. Also, $C_{0,2}$ are zero- and two-form potentials, $B_2$ 
 	is the Kalb-Ramond  field, and  $ \Omega(z_a)$ the holomorphic three-form which depends on the complex structure moduli denoted hereafter with $z_a$.
 	 Supersymmetric conditions imposed on ${\cal W}_0$ stabilize the dilaton and -in principle- the complex structure moduli (for recent activity on this issue see section 3).  However, the K\"ahler structure ones, $\rho_k$,  do not appear
 	 in the tree level superpotential and thus their values remain undetermined. Moreover,  as it is described in well known non-renormalization 
 	 theorems for fluxed superpotentials in string theory~\cite{Burgess:2005jx},  ${\cal W}_0$  cannot  receive perturbative corrections to any order and the  only possible contributions are of non-perturbative type. 
 	  Regarding the  K\"ahler moduli fields, non-perturbative corrections may arise from several sources including 
 	  $D3/\overline{D3}$ branes wrapping four-cycles  and gaugino condensation on $D7$-brane stacks. 
   In general all 
 	 three  moduli of the present construction 
 	 may contribute, hence the generic form of superpotential  is
 	\be
 	\label{Wnp3}
 	\begin{split}
	{\cal W}&= {\cal W}_0+{\cal W}_{np}
	\\
	&={\cal W}_0+ \sum_{k=1}^3 A_k e^{i a_k\rho_k}~.
 	\end{split}  
 	\ee 
 	In~(\ref{Wnp3}) the flux induced part	${\cal W}_0$ will be considered  constant evaluated by the formula given in~(\ref{GVW}).
     The coefficients  $A_i$ are functions of $z_a$, and $a_k$ are 
 	small parameters which in the case of gaugino condensation
 	take the form $a_k=\frac{2\pi}{N_k}$, with $N_k$ the rank of 
 	the corresponding gauge group of the $D7$ brane stack. 
 	In~\cite{Basiouris:2020jgp} the simplest scenario of only one  K\"ahler  modulus field (say $\rho_1$) was considered  to have non-vanishing non-perturbative (NP) contributions
     so that the superpotential ~(\ref{Wnp3}) reduces to ${\cal W}= {\cal W}_0 + A e^{-a\rho_1}$.

     The second important ingredient is the     K\"ahler potential  which depends 
     logarithmically on   the K\"ahler fields $\rho_i$, the complex structure moduli $z_a$ through the 3-form $\Omega(z_a)$, and the axion-dilaton field $S$. At tree level this reads
     \begin{eqnarray} 
     	{\cal K}_0&=&-2 \ln\left(\sqrt{-\frac{i}{8}({\rho_1-\bar{\rho_1}})({\rho_2-\bar{\rho_2}})({\rho_3-\bar{\rho_3}})}\right)
     	- \ln(-i({ S-\bar{S}}))-\ln(-i\int{\Omega}\wedge {\bar{\Omega}})\nn\\
     	&=&-2 \ln\left({\cal V}\right)
     	- \ln(-i({ S-\bar{S}}))-\ln(-i\int{\Omega}\wedge {\bar{\Omega}})
     	~,\label{KahlerP}
     \end{eqnarray}
     where in the second line the formulae~(\ref{3rhos}) and~(\ref{svol}) have been implemented. 
     
     \noindent 
    It has been shown~\cite{Antoniadis:2019rkh}  that the argument of the logarithmic term $-2\log{\cal V}$ of the K\"ahler potential~(\ref{KahlerP})  receives perturbative logarithmic corrections of the form $\delta_{\cal V}=\xi+\eta \log({\cal V})$, where $\eta$ is a negative constant  ($|\eta|\sim O(1)$) and $\xi$ depends on the Euler characteristic of the compactification manifold. As a result, the induced effective scalar potential   includes logarithmic terms which are  expressed in terms of the internal volume modulus ${\cal V}$.  In addition, the total F-term potential involves other terms with  different (power-law) dependence on ${\cal V}$.  In effect,  the structure of the potential is reminiscent of the Coleman-Weinberg mechanism~\cite{Coleman:1973jx} and  its minimum is found at finite values of the volume modulus.\\
     Furthermore,  additional   D-terms of the form ${d_i}/{\tau_i^3},\,i=1,2,3$  associated with the universal $U(1)$ factors of the $D7$ brane stacks 
     can be  included. 
 When all these components are taken into account, in the large volume limit 
 the scalar potential can be  approximated as follows (see Appendix for details)

\begin{equation}
V_{\rm eff}\approx (\epsilon \mathcal{W}_0)^2\left(2\dfrac{\xi+2\eta(\log(\mathcal{V})}{\mathcal{V}^3}-\dfrac{1}{\mathcal{V}^2}\right)+\dfrac{3{d}}{\mathcal{V}^2}~.\label{Veff1}
\end{equation}
The parameter $\epsilon =\frac{2}{1+(2a_1\tau_1)^{-1}}$ has been  considered   in the large volume limit  $a_1\tau_1\gg 1$ to be  $\eps\approx 2$, and $d$   is defined as the product  ${d}=(d_1d_2d_3)^{1/3}$. 
 Hence, the two (unspecified yet) constants   $(\epsilon  \mathcal{W}_0)$ and $d$  in (\ref{Veff1})   multiply  the F- and D-term parts  respectively.   
Comparing  the effective potential~(\ref{Veff1}) with that obtained in~\cite{Antoniadis:2018hqy} -where only perturbative 
corrections are taken into account- 
it is observed that the latter contains an additional  term  $\propto -\frac{1}{{\cal V}^2}$. This term, however, can be absorbed in
the D-part of $V_{\rm eff}$ under a redefinition of the constant $d\to \tilde d= d-(\epsilon \mathcal{W}_0)^2/3$.  As it has been 
demonstrated in~\cite{Basiouris:2020jgp}, for given $\eta$ and $\xi$, as long as
the ratio of the D- and F-term coefficients  $r=\frac{\tilde d}{(\epsilon{\cal W}_0)^2}$
is bounded in the narrow region, $r\in [\frac{1}{12}+\frac{|\eta|}{3{\cal V}_{min}}\,,\, \frac{1}{12}+\frac{7|\eta|}{8{\cal V}_{min}}]$ a dS minimum is attainable  and all three K\"ahler moduli are stabilized.

 	\subsection{The superpotential with two non-perturbative terms  }
 	
 As it has been emphasized  above, the case of a single non-perturbative term in the superpotential ensures
 moduli stabilization and the existence of a dS
 vacuum as long as perturbative logarithmic corrections 
 and D-term contributions are included. However,  due to the  variety of compactification manifolds  the case 
 of non-perturbative contributions from  more than one moduli is
 a more likely scenario. Given the fact that the shape of 
 the potential near the minimum is of particular importance for
 cosmological applications, (and in particular,  inflation) it is worth 
 exploring  more involved situations. 
 The purpose of  the present  work is to extend  the previous analysis, 
  within the same  geometric configuration of three  $D7$ branes,	and examine the implications  when   the flux induced  superpotential  
 	$ {\cal W}_0$,  receives 
 	non-perturbative corrections from  two  K\"ahler moduli, $\rho_1$ and $ \rho_2$. 
In this case the superpotential    takes the form
 	\be  
 	{\cal W}= {\cal W}_0+ A e^{i a \rho_1}+ B e^{i b \rho_2}~,\;{\rm with} \; a>0\, {\rm and}\, b>0~.\label{Wnp2}
 	\ee
Cases with two exponentials  capture many new features and  have been discussed in the literature in particular constructions. Generically, one would anticipate  non-perturbative corrections  in every direction of the moduli space. It is possible, however,  that world volume fluxes can in principle modify the effective action through lifting certain fermionic zero-modes, which could prohibit a particular contribution in the superpotential~\cite{Bianchi:2011qh}. In the present analysis
this mechanism is  considered to eliminate from ${\cal W}_{np}$ the exponential term involving the modulus $\rho_3$ . In a  recent work  for example, the two exponential terms are generated by E3 instantons wrapping appropriate singularities~\cite{Cicoli:2021dhg}. Also, the racetrack  form~\cite{Blanco-Pillado:2004aap}
 suitable for cosmological applications  could be considered as  a particular case when both   exponents of~(\ref{Wnp2})
 involve   the  same modulus, i.e., when $\rho_2$ in replaced with $\rho_1$ in the second exponential.  
In general,  two or more exponential terms imply a richer structure for the shape of $V_{\rm eff}$ which could exhibit  saddle points 
between different vacua of the theory, so that successful types of inflationary scenarios  can be realized~\cite{Linde:2007fr}. 
Despite the vast literature devoted on such issues,	the combined effects of (\ref{Wnp2})   with perturbative 
	logarithmic corrections to the K\"ahler potential have not been investigated so far. These  ingredients 
	are a generic feature  of the effective theories derived 
	from the 10-dimensional superstring action and thence it is the main
	subject of the subsequent analysis.

 	Starting with the superpotential of the effective theory, while recalling 
 	that in general the coefficients $A$ and $B$ depend on the complex structure moduli $z_a$, it can be readily inferred that supersymmetric conditions imposed on~(\ref{Wnp2}) can fix the axion-dilaton field and $z_a$. Since $z_a$ receive masses of
 	order $m_{z_a}\sim \alpha'/R^3$ whilst K\"ahler masses arise through the non-perturbative corrections, the former are expected to be much larger and can be integrated out~\cite{Douglas:2006es}.  Thence the last two logarithmic factors of~(\ref{KahlerP}) involving  these fields can be treated as constants.

 	\subsection{The K\"ahler potential  }
 	
 When various types of  perturbative corrections  are included, the no scale structure of the classical  K\"ahler potential~(\ref{KahlerP})  is no longer preserved.  Of crucial importance in the present study, are the  higher derivative terms of the 10-d string action, which are responsible for multigraviton scattering.  
 	Indeed, it was found in~\cite{Antoniadis:2019rkh} that in combination with intersecting  D7-branes and $O7$ planes included in the present geometric configuration, 
  volume dependent logarithmic corrections  are induced in the K\"ahler potential. 
 	 For the sake of completeness the  basic steps of their    derivation are resumed in the following.
  
  In the 10-d action the terms next to leading order  in curvature $\mathcal{R}$  are proportional to $\mathcal{R}^4$~\cite{Antoniadis:1997eg},\cite{Antoniadis:2003sw},\cite{Kiritsis:1997em}:
 	 \begin{eqnarray} 
 	 	{\cal S }&{ \supset }&\frac{1}{(2\pi)^7 \alpha'^4}  \int\limits_{ M_{10}} e^{-2\phi_{10}} {\cal R}_{(10)}  + \frac{6}{(2\pi)^7 \alpha'} \int\limits_{M_{10}} (2\zeta(3) e^{-2\phi_{10}} + 4\zeta(2) )  {\cal R}_{(4)} ^4 \wedge e^2\,,
 	 \end{eqnarray}
 	 where ${\cal R}_{10}, M_{10}$  and $\phi_{10}$ 
 	 are the $10$-dimensional Ricci scalar,  manifold and dilaton respectively.
 	 Compactifying six dimensions on a CY manifold ${\cal X}_6$, so that $M_{10}={ M}_4\times {\cal X}_6$ with ${ M}_4$ the Minkowski space,
 	 while  taking into account the tree-level and  one-loop generated EH terms,  the ten-dimensional action reduces to  
 	 \begin{eqnarray} 
 	 	{\cal S }_{\rm grav}&=& \frac{1}{(2\pi)^7 \alpha'^4} \int\limits_{M_{4} \times {{\cal X}_6}} e^{-2\phi_{10}} {\cal R}_{(10)} + \frac{\chi}{(2\pi)^4 \alpha'} \int\limits_{M_{4}} \left(2\zeta(3) e^{-2\phi_{10}}  + 4\zeta(2) \right) {\cal R}_{(4)}\,,    
 	 	\label{IIBAction} 
 	 \end{eqnarray} 
 	 where $\chi$ is the Euler characteristic defined as
 	 \be 
 	 \chi= 	\frac{3}{4\pi^3}\int\limits_{{\cal X}_6}{\cal R}\wedge {\cal R} {\cal\wedge R}~\cdot 
 	 \label{Eulerx}
 	\ee
 	 This way a localized EH term   ${\cal R}_{(4)}$ is generated, 
 	 with a coefficient proportional to the Euler characteristic $\chi$,  whose existence   is possible only in four dimensions. 
 	In the case  of type	IIB string theory compactified on the $T^6/Z_N$ which will be considered in this work, non-zero $\chi$ values are   concentrated at the fixed points of the orbifold group. Moreover,   localized vertices associated with the induced EH term, emit gravitons and Kaluza-Klein (KK) excitations in the 6-dimensional space. A useful notion related to this  mechanism is the localization width $w$ 
 	 which can be estimated by computing the graviton scattering involving two massless gravitons  and one KK excitation.  It is found that~\cite{Antoniadis:2002tr}
 	 \begin{eqnarray}  
 	 	w \approx \frac{l_s}{\sqrt N}~,
 	 	\label{Width}
 	 \end{eqnarray} 
 	 with $l_s=\sqrt{\alpha'}$ the fundamental string length and the integer $N$ (which is arbitrary in the non-compact limit) is associated with the Euler  characteristic  $\chi\sim N$. 
 	 
 	
Due to  $D7$ branes and orientifold $O7$ planes of the assumed geometric configuration, there can be an exchange of KK-modes between the localized gravity position identified with some orbifold fixed-point, and the distinct D7-branes.  Because the latter  occupy four internal dimensions, KK-modes transmitted towards each one of them propagate  in a  two-dimensional bulk transverse to $D7$.   This way,  logarithmic contributions  are induced depending on the size $R_{\bot}$ of the two-dimensional space transverse to $D7$.
 Thus, in addition to
the one loop contribution, there are also logarithmic corrections.
Referring to the relevant works for the details~\cite{Antoniadis:2002tr}, one finds that the corresponding term of the effective action becomes: 
 	 \begin{eqnarray}
 	 	\frac{4\zeta(2)}{(2\pi)^3}\chi \int_{M_4} \left(1-\sum_k e^{2\phi}T_k \ln(R^k_{\bot}/w)\right)\,{\cal R}_{(4)}\,,
 	 	\label{allcor}
 	 \end{eqnarray}
 	  where $T_k$ is the tension of the $D7_k$ brane and $R^k_{\bot}$ the 
 	 size of the  2-d space  transverse to the corresponding $D7_k$. 
	From~(\ref{allcor})  it can be deduced that the quantum corrections   are of the form
 	\be 
 	{\delta} =\xi+ \sum_{j=1}^3{ \eta_j} \ln ({ \tau_j})~.
 	\ee  
 	The constant $\xi$ is proportional to the Euler characteristic $\chi$ and 
 	in the case of orbifolds  for example, is given by  $\xi=  - \frac{\pi^2g_s^2}{12}\,\chi$.
 		
Incorporating these  quantum corrections in the  K\"ahler potential
while substituting   $\rho_k-\bar \rho_k=2 i \tau_k$ from (\ref{3rhos})
the corrected K\"ahler potential (\ref{KahlerP}) becomes
 		\begin{eqnarray} 
 		{\cal K}&=&-2 \ln\left(\sqrt{\tau_1\tau_2\tau_3}+\xi+\sum_{j=1}^3{ \eta_j} \ln ({ \tau_j})\right)+\cdots
 		~,\label{KahlerCorrected}
 	\end{eqnarray}
 where $\cdots$ stand for terms involving complex structure
 moduli and the axion-dilaton, as in~(\ref{KahlerP}). 
 	Assuming for simplicity 
 	that all $D7$ branes have the same tension proportional to $T_0$,  
 	the coefficients $\eta_j$ take a common value $\eta_j\equiv \eta =-\frac{1}{2} g_sT_0\xi $, and  Eq.(\ref{KahlerCorrected}) yields
 
 	\begin{equation}
 		{ \cal  K}=
 		-2\ln\left({\cal V}+{\xi}+ {\eta} \ln {\cal V}\right)+\cdots ~. \label{QCK}
 	\end{equation}

 	\subsection{The supersymmetric conditions}
\begin{figure}[H]
 		\centering
 		\includegraphics[scale=0.65]{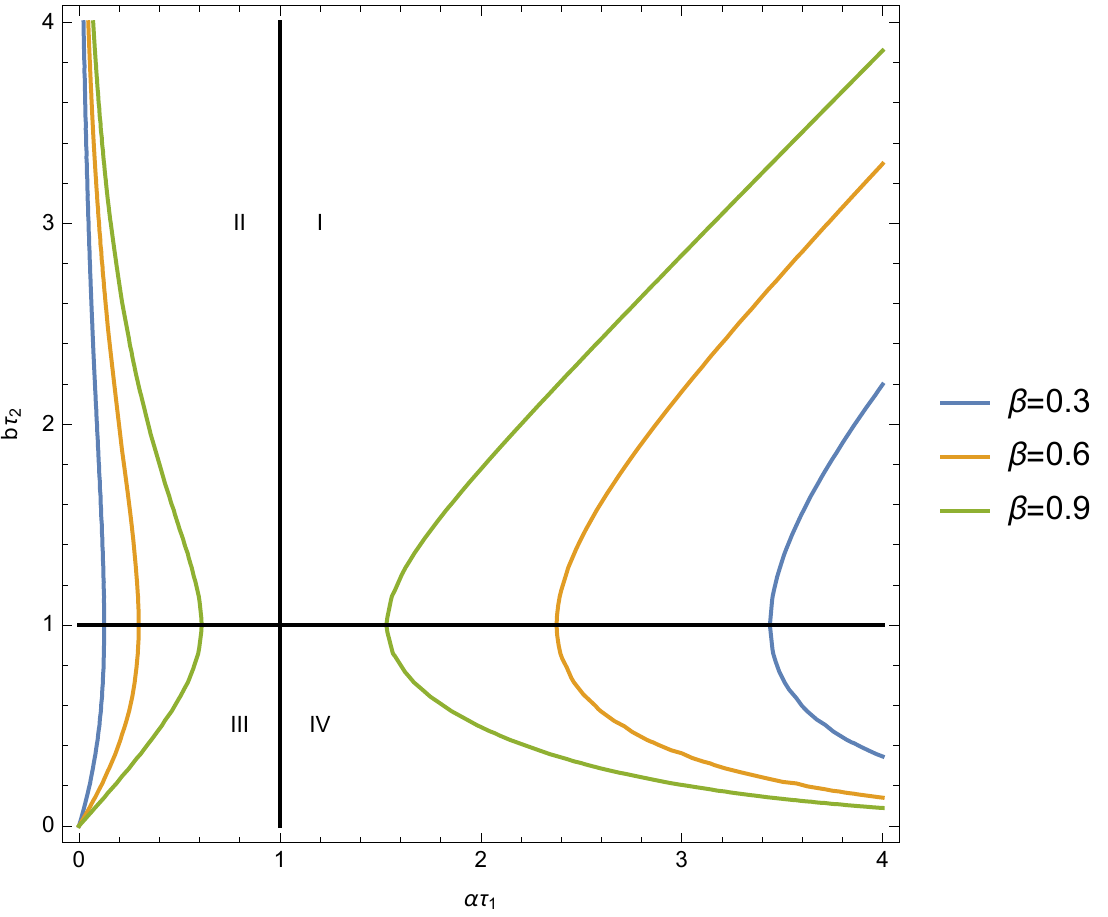}\;\includegraphics[scale=0.65]{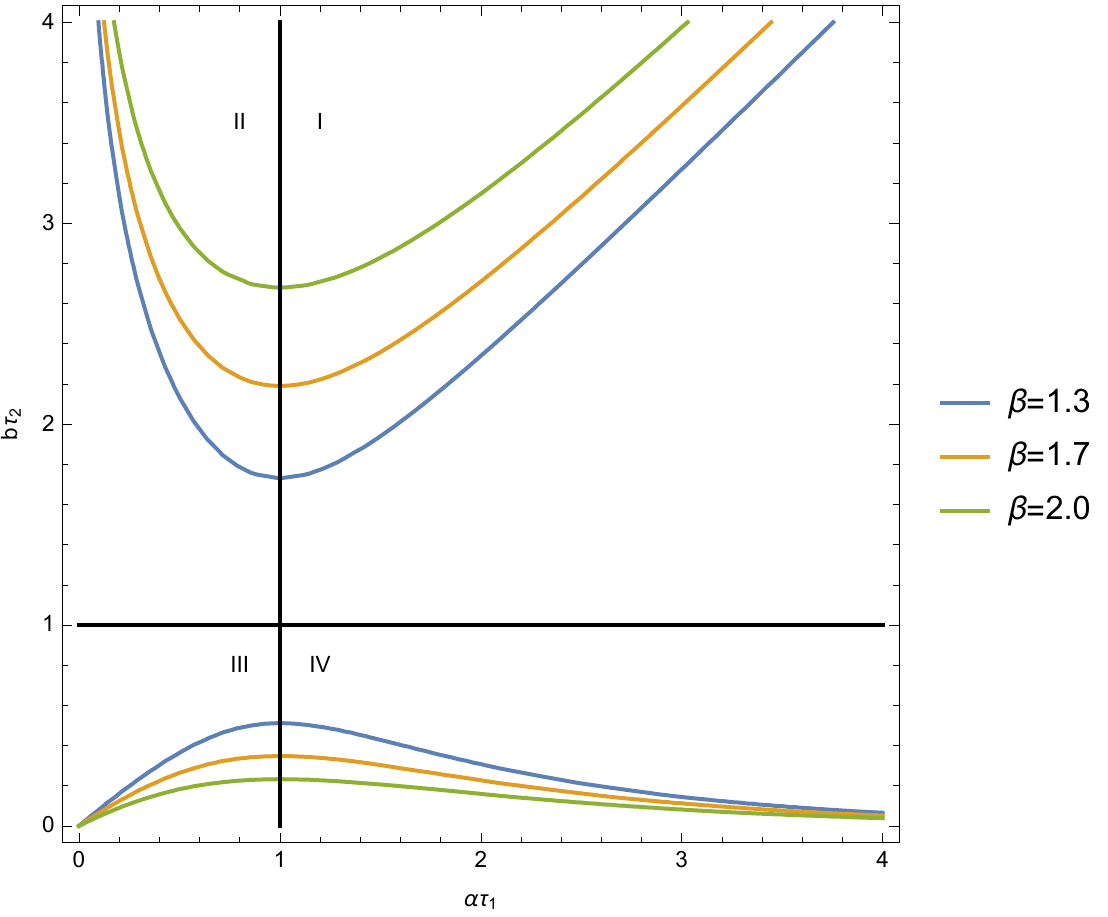}
 		\caption{Graphical  solution of Eq~\eqref{contourequ} for various values of the parameter $\beta=B/A$ defined in~(\ref{ratios}).
 			 The left panel shows curves for three 	values of $\beta <1$ and the right panel for $\beta>1$. } 
 		\label{contourst1t2}
 	\end{figure}

 	In the present setup, the contribution of the moduli $\rho_1,\rho_2$ in the  superpotential enters through  the non-perturbative corrections, and thus, 
 	the appropriate flatness conditions must be imposed.   The latter imply  the vanishing of the corresponding covariant
 	 derivatives $D_{\rho_i}W= \partial_{\rho_i} {\cal W}+{\cal W}\partial_{\rho_i} {\cal K}$. Introducing the expansions 
 	with respect to   $\eta/{\cal V}$ and $\xi/{\cal V}$ in the large volume limit, 
 	it is readily found that
 	\begin{align}
 		D_{\rho_1}\mathcal{W}|_{\rho_1=i\tau_1}^{\rho_2=i\tau_2}&=-\dfrac{A(e^{-a \tau_1}(1+2a \tau_1)+\beta e^{-b \tau_2}+\gamma)}{2\tau_1}+\mathcal{O}(\eta,\xi)=0,\label{flatW1}\\
 		D_{\rho_2}\mathcal{W}|_{\rho_1=i\tau_1}^{\rho_2=i\tau_2}&=-\dfrac{A(e^{-a \tau_1}+e^{-b\tau_2}(1+2b\tau_2)\beta+\gamma)}{2\tau_2}+\mathcal{O}(\eta,\xi)=0~,\label{flatW20}
 	\end{align}
 	where $ \beta, \gamma,$ stand for the following ratios :
 	\begin{equation}
 		\beta=\dfrac{B}{A},\; \gamma=\dfrac{\mathcal{W}_0}{A}~.
 		\label{ratios}
 	\end{equation}
 If some reasonable assumptions concerning  the various flux parameters and the range of moduli fields are made,  the solutions of 
 the above transcendental equations can be expressed in closed form with good accuracy, in terms of known functions. 
 A possible  choice of	the approximations  can be better perceptible as follows: The two
 	equations (\ref{flatW1}) and (\ref{flatW20}) are combined to 
 	give 
 	\be 
 	a \tau_1\,e^{-a \tau_1}=
 	\beta\,b\tau_2\,e^{-b\tau_2}~. \label{contourequ}
 	\ee 
 	Since $a,b$ are positive constants, it turns out that $\beta>0$, while real  solutions of  (\ref{flatW1},\ref{flatW20}) exist as long as $\gamma<0$.
 	The equation~(\ref{contourequ}) is plotted in figure (\ref{contourst1t2}) for several values of  $\beta$ in the  parametric space defined by  the pair ($a\tau_1, b\tau_2$). The curves of the left panel correspond to values $\beta <1$ and the ones on the right, to    $\beta >1$. (For $\beta=1$ a trivial solution exists $a\tau_1=b\tau_2$ represented by the diagonal, not shown in the figure). The parametric space has
 	been split into four regions $I,II,III, IV$ with respect to the ranges
 	of  $a\tau_1$ and $b\tau_2$.  Region $I$ corresponds to large 
 	values of $a\tau_1, b\tau_2$ and thus,  both terms of the non-perturbative
 	contributions in (\ref{Wnp2}) are suppressed. In general,
 	in the large volume regime,   perturbative logarithmic 
 	corrections are expected to prevail.

 In the opposite limit, region $III$ corresponds to small
 	values of $a\tau_1, b\tau_2$, and both  NP-contributions become sizable,
 	however, in this case large ${\cal V}$   requires 
 	$\tau_3$-values much bigger  than   $\tau_1, \tau_2$. 
 	A drawback of this region is that non-perturbative 
 	corrections correspond to the large coupling regime and as such they are not fully controllable. Nevertheless, for the sake of completeness a short analysis 
 	will be presented in a subsequent section.
 	
 Finally, the regions $II$ and $IV$,  for typical values of the gaugino condensation  parameters
 	$a= \frac{2\pi}{M} \sim b= \frac{2\pi}{N} $, can be associated with cases where there could be   a milder
 	hierarchy between the moduli  fields $\tau_{1,2,3}$. Then, at least one NP-term in~(\ref{Wnp2}) could make
 	significant contribution to the superpotential and it would be 
 	interesting to investigate its implications.

 	\section{The F-term scalar potential}
 		The fundamental quantity for the study of the effective field theory vacuum is  the scalar potential $V_{\rm eff}$. In the present study this is comprised
 	of the F- and D-term, written as $V_{\rm eff}=V_F+V_D$ (in a self explanatory notation), and will be examined in detail in the large volume regime. This section
 	deals with the  contribution of the F-term   $V_F$. 
 	
 	The  present study will proceed with the  investigation  of the properties of $V_{\rm eff}$ in reasonable parts  of  the  regions defined in figure~\ref{contourst1t2}, that is, regions with $a\tau_{1}\ll 1$ and $b\tau_{2}\ll 1$ will be excluded from the analysis. 
     In the  present section the F-term scalar potential will be analyzed 
 	and as a first approach, the   restriction 
 	\begin{equation}
 		\beta e^{-b\tau_2}\ll |\gamma|\, \leftrightarrow \, B  e^{-b\tau_2}\ll |{\cal W}_0|~, \label{approx}
 	\end{equation}
 will be imposed which entails a  non-perturbative part $Be^{-b\tau_2}$ much smaller   than  the flux induced  tree-level superpotential $|{\cal W}_0|$. It should be noted  that in the large volume regime  small 
 fluxes discussed in recent works~\cite{Demirtas:2019sip,Alvarez-Garcia:2020pxd,Broeckel:2021uty,Honma:2021klo}, are not excluded by the assumption imposed above. 	For example, for ${\cal W}_0\sim 10^{-8}$, condition~(\ref{approx}) is satisfied\footnote{  
 	Considering the recent activity for the quest of vacua with exponentially small $\mathcal{W}_0$, it would be worth  commenting on this parametric region. According to~\cite{Denef:2004ze}, the plethora of flux vacua could be described as a statistical ensemble where the value of  $\mathcal{W}_0$ plays a significant r\^ole. Models with $\overline{D3}$ uplift, such as~\cite{Kachru:2003aw},  are based on the conifold geometry for the D-brane configurations~\cite{Giddings:2001yu,Choi:2004sx}, since the dilaton and the CS moduli are parametrically heavier than the K{\"a}hler fields and could be effectively integrated out. A large amount of CS moduli  (which is the case for the most well studied CY manifolds)  requires   big $D_3$ charges in order to satisfy the tadpole cancellation. Consequently, this implies small values for $\mathcal{W}_0$ at the weak coupling regime as it is also predicted by the statistical analysis.}, for $\beta\sim O(1)$ and $b\tau_2>20$.
 
 	 As it will be seen in the subsequent analysis, in this limiting case  it is  possible to present sufficiently accurate analytic formulae for the flatness solutions and achieve  a compact form of $V_{\rm eff}$.   A different approach  where this condition is relaxed will be presented in  a subsequent section.

 From~(\ref{contourequ}) the first term of (\ref{approx}) is
 $	\beta\,e^{-b\tau_2}= \frac{	a \tau_1}{b\tau_2}\,e^{-a \tau_1}$. Hence, the approximation 
  is valid  for  small fluxes associated with 
 	the coefficient $B$  and/or  large hierarchies $b\tau_2\gg a \tau_1$.
 	Thus, the  focus  of the analysis in the present section  will be on the  appropriate sections of the regions $I$ and $II$ where  the hierarchy $a\tau_1\ll b \tau_2$ 
 	holds (a similar  analysis for region $IV$ is appropriate for $a\tau_1\gg b \tau_2$). The case of region $III$ will be analyzed using
 	a different parametrization. 
 	
 	 In addition, 
 	the energy scale and the coefficients $a, b$ related to gaugino condensations  on each brane can differ.
 	Under these assumptions, the equations (\ref{flatW1},\ref{flatW20}) reduce to:
 	\be 
 	\label{flateq2}
 	\begin{split}
 		D_{\rho_1}\mathcal{W}|_{\rho_1=i\tau_1}^{\rho_2=i\tau_2}&=-A\dfrac{e^{-a \tau_1}(1+2a \tau_1)+\gamma}{2\tau_1}\approx 0,\\
 		D_{\rho_2}\mathcal{W}|_{\rho_1=i\tau_1}^{\rho_2=i\tau_2}&=-A\dfrac{e^{-a \tau_1}+ 2b\tau_2\beta\,e^{-b\tau_2}+\gamma}{2\tau_2}\approx 0~.
 	\end{split}
 \ee 
 	It is convenient to  solve the above with respect to the moduli fields   $\tau_1, \tau_2$. Defining the new 
 	variables $w,u$ 
 	\begin{align}
 	w	=-\dfrac{1+2a\tau_1}{2},&\;\;
 	u=-b	\tau_2~,\label{sol12}
 	\end{align}
 the solutions are expressed as follows~\footnote{For example, 
 	 the two equations imply\\ $e^{-a\tau_1} (1+2a\tau_1) =-\gamma\,\Rightarrow \,2 w e^{-a\tau_1}=\gamma$ or $w e^{w}=\frac{\gamma}{2\sqrt{e}}$ etc.}
 	\be 
 	\label{Lambert}
 	\begin{split}
 		w&\equiv  w(\gamma)=W(\frac{\gamma}{2\sqrt{e}}),\\
 		u&\equiv  u(\gamma)=W\left(\frac{\sqrt{e}\;e^{w}+\gamma}{2\beta}\right)\\
 		&\equiv W\left(\frac{\gamma}{\beta}\frac{1+2w}{4w}\right)~.
 	\end{split}
 	\ee 
 In the above solution,  $W$  stands for either of  the two branches $W_{0},W_{-1}$ of the Lambert-W function. For large $\tau_2$ values however,  the function $W$ in  (\ref{Lambert}) should be identified with the branch 
 	 $W_{-1}$.
 	For later convenience, the following parameters are also introduced:
 	\be 
 	\eps\, =\,\frac{1+2 w}{w},\;\; \tilde\eps\,=\,\frac{\eps}u~.\label{tildeps}
 	\ee
 	The restriction to real values of the two branches $W_0, W_{-1}$ imposes the bounds  on the various new parameters   shown in  Table 1.  The approximation~(\ref{approx})  is valid only for regions $I$ and $II$ where $u\equiv -b\tau_2 <-1$.
 	\begin{table}[H]
 		\begin{center}  \small%
 			\begin{tabular}{|P{2cm}|P{2cm}|P{2cm}|P{2cm}|P{2cm}| P{2cm}|}
 				\hline
 				& $\gamma$ & $\beta$ & $w$ & $u$&$\tilde{\epsilon}$ \\
 				\hline
 				$I$ & $(-\frac{2}{\sqrt{e}},0)$ & $(0,\infty)$ & $(-\infty,-\frac{3}{2})$ & $(-\infty,-1)$ & $(-2,0)$\\
 				\hline
 				$II$ & $(-\frac{2}{\sqrt{e}},-1)$ & $(0,\infty)$ & $(-1,-\frac{1}{2})$ & $(-\infty,-1)$ & $(-1,0)$\\
 				\hline
 				$III$ & $(-\frac{2}{\sqrt{e}},-1)$ & $(0,\infty)$ & $(-1,-\frac{1}{2})$ & $(-1,0)$ & $(-\infty,0)$\\
 				\hline
 				$IV$ & $(-\frac{2}{\sqrt{e}},0)$ & $(0,\infty)$ & $(-\infty,-\frac{3}{2})$ & $(-1,0)$ & $(-\infty,0)$\\
 				\hline
 			\end{tabular}
 		\end{center}
 		\caption{Limiting values of different parameters for each one of the regions depicted in Figure~\ref{contourst1t2}.}
 	\end{table}

Equipped with the above formulae and the assumption  that the complex structure moduli are already fixed at large values (and therefore decoupled from the spectrum), it is possible to 
compute the F-term potential through the well-known supergavity
formula
\be
V_F= e^{\cal K}\left(\sum_{I,J} {\cal D}_I{\cal W}{\cal K}^{I\bar J}{\cal D}_{\bar J}{\cal W}-3 |{\cal W}|^2\right)
	~\,, \label{VFKahler}
\ee

 	Before computing  the K\"ahler moduli contribution to $V_F$, which is the main focus of this work, a few more  remarks regarding   the complex structure (CS) moduli  will be made with emphasis on recent progress~\cite{Grimm:2018cpv,Grimm:2019ixq,Braun:2020jrx,Marchesano:2021gyv}.    As can be seen from ~(\ref{KahlerP}), the CS moduli enter  in the K\"ahler potential through the logarithmic  term $K_{cs}=-\log(-i\int \Omega\wedge \bar \Omega)$. A detailed treatment of the minimization with respect to CS moduli requires knowledge of the periods $\Pi^I$ 
 	of the holomorphic (3,0)-form $\Omega$.  The latter can be expanded in terms of its  periods as  $\Omega =\Pi^I \gamma_I$ where $\gamma_I$ is  a suitable real basis $\gamma_I=1,\dots, 2 h^{2,1}+2$ and $ h^{2,1}= {\rm dim}[H^{2,1}(Y_3)]$  is the dimension of the CS moduli space. Then, the associated part of the K\"ahler potential is expressed as  $K_{cs}=-\log(\Pi^I\eta_{IJ}\bar\Pi^J)$, where $\eta_{IJ}$ defines the integral form of  the antisymmetric matrix~\cite{Grimm:2018cpv} $\eta_{IJ}=-\int_{Y_3}\gamma_I\wedge \gamma_J$ over the Calabi-Yau threeforld $Y_3$.
 	The  periods ${\bf \Pi}\equiv \{\Pi_1,\dots, \Pi^{2h^{2,1}+2}\}$  are complicated holomorphic functions subject to monodromies when going around divisors comprising  the discriminant locus, i.e., the points at which the CY manifold becomes singular~\cite{Grimm:2018cpv}.
 	 In the limit of large complex structure, however, it is possible using the nilpotent orbit theorem~\cite{Schmid1973} to express them in terms of 
 		a simple formula  modulo exponentially suppressed terms. Indeed, introducing local coordinates $z^j$ which determine the divisors by $z^j=0$,  
 	according to the nilpotent orbit theorem the periods are approximated as follows
 		\be 
 	 {\bf \Pi}_{nil}= e^{\sum_j-t^jN_j}\left( {\bf a_0}+ {\cal O}(e^{2\pi i t})\right)~.
 		\ee 
 	In the above formula, the summation index $j$ runs over all divisors,  $N_j$ are nilpotent matrices encoding the singular behavior,  $t^j=x^j+i y^j=\frac{1}{2\pi i}\log(z^j)$, and  ${\bf a}_0(\zeta)$ depends on coordinates $\zeta^k$ other that $z^j$ defined above. 
 	In~\cite{Marchesano:2021gyv} the large complex structure approach has been  implemented to discuss F-theory vacua with respect to the CS moduli fields.  The form of the periods of $\Omega$ at large complex structure where extracted using mirror symmetry between type IIA and IIB/F-theories.   It was found that the scalar potential acquires a simple form $V_{cs}=\frac 12 {\cal Z}^{AB}\varrho_A\varrho_B$ where $\varrho_{A,B}$ represent
 	monodromy invariant functions of axion parts and fluxes yielding a Minkowski vacuum~\footnote{According to recent works~\cite{Bena:2020xrh,Braun:2020jrx} it is debatable  whether  stabilization of all complex structure moduli is possible  since magnetic fluxes providing them with masses  are constrained by tadpole cancellation conditions. However, specific counterexamples have been designed~\cite{Marchesano:2021gyv} in the context of F-theory where the tadpole conjecture can be evaded. This issue is still open and under investigation.}.  In the present work, as has been already emphasized, the subsequent analysis	will proceed  with the assumption  that the complex structure moduli are already fixed at large values (and therefore decoupled from the spectrum). 

Next, concentrating  on the K\"ahler moduli contributions, it is observed that 
due to the simultaneous presence of non-perturbative exponential contributions  in~(\ref{Wnp3})    and the logarithmic corrections in the K\"ahler potential~(\ref{KahlerCorrected}) the F-part  $V_F$ of the scalar potential  is a very complicated function.
 	Nevertheless, in  the large volume limit, and under the assumptions
 	$|\eta|\lesssim 1$ and $\xi\ll {\cal V}$, $V_F$  receives an approximated closed 
 	form which is  sufficiently  accurate to  capture all the essential features regarding the
 	moduli stabilization and the existence of dS vacua.

Formally the $V_F$ term comprises of three parts, the pure perturbative and  non-perturbative parts and a term which is a mixing of both.  Before presenting the total $V_F$, 
it is useful to examine separately the form of the perturbative and non-perturbative parts. For example,	
implementing the expansion with respect to $\eta$ and $\xi/{\cal V}$ the perturbative part receives
the following simplified form (details can be found in the Appendix)
 	\be 
 	\begin{split}
 		V_F^{(p)}&\approx 
  	\frac{3}{2} {\cal W}_0^2 \frac{\xi +2 \eta  \log ({\cal V})}{ {\cal V}^3}+{\cal O}(\frac{1}{ {\cal V}^4})~.\label{Fpapprox}
 \end{split}
 	\ee 
 	From this simplified form of the perturbative part~(\ref{Fpapprox}) it is  observed that the numerator 
 	consists of two terms of different volume dependence. For 
 	$\eta<0$ and $\xi>0$ in particular $	V_F^{(p)}$ 
 	acquires a minimum at ${\cal V}_0=e^{\frac{1}{3}+\frac{ \xi }{2 |\eta| }}$, however the value of the potential at the minimum is negative, $(V_F^{(p)})_{min}=\frac{2}{3} \eta  e^{\frac{3 \xi }{2 \eta }-1}<0$, i.e., it defines an Anti de Sitter (AdS) vacuum. 
 	
 \noindent The pure non-perturbative part $	V_F^{(np)}$
 	becomes (for the derivation see Appendix)
 	\be 	
 	V_F^{(np)}=   -{\cal W}_0^2 \frac{(u+1) (2 w+1)^2}{2 u w^2 {\cal V}^2}\equiv -(\tilde\eps  {\cal W}_0)^2\,\frac{u (u+1)}{2  {\cal V}^2}~.\label{FNPpure}
 	\ee 
 	Remarkably, this term has a volume dependence $\propto  \frac{1}{{\cal V}^2}$ which is exactly the dependence 
 	of the D-term uplift in~(\ref{Veff1}).
 	For the regions $I,II$ where the approximation is valid,	 however, because $u (1+u)>0$ the contribution of this term 
 	is negative and deepens the AdS vacuum.\footnote{ Nonetheless, it will be seen that this term  has the same power-law volume   dependence with the positive D-term contributions $d/{\cal V}^2$ and can be compensated by appropriate values of the  parameter $d$.}

	The full F-part of  the scalar potential comprising all those three parts can 
 	 be  written in a simple form using the substitutions (\ref{wtaurelationsA}) given in the Appendix and the exact relation 
 	 (\ref{contourequ}). These manipulations yield 
 	\ba 
 	V_F&\approx & (\tilde\eps  {\cal W}_0)^2\left( -\frac{u (u+1)}{2  {\cal V}^2}+
 	\frac{ (2 u+1) (14 u+3) (\xi+2 \eta) \log ({\cal V})-24 \eta }{32 {\cal V}^3 }\right.\nn
 	\\
 	&&\qquad\qquad\qquad\qquad\qquad\qquad \qquad  \left.+\eta  \xi
 	\frac{48 u-  \left(68 u^2+60 u+9\right) \log{\cal V}}{32 {\cal V}^4}\right)~.\label{Veff0}
 	\ea 
 	It is again emphasized that this form is valid for the regions $I,II$ and cannot be used to describe the physics for regions $III$ and $IV$. 
 	In the large volume case where the term  $\propto  \frac{1}{{\cal V}^4}$ can be safely ignored, the minimum of the potential   
 	for the volume modulus can be  found analytically. 
  Setting	the first derivative equal to zero and solving, the volume at the minimum is found to be 
 	\be 
 	{\cal V}_{min}=-{\eta p(u)}\, W_0\left(-\frac{1}{\eta p(u)}{e^{q(u)-\frac{\xi }{2 \eta }}}\right)~,
 	\label{VolVefmin}
 	\ee 
 	where, for the subsequent analysis the following  convenient parametrization has been introduced 
 	\be\begin{split}
 		p(u)&=\frac{3}{16}\,\frac{   (2 u+1) (14 u+3)}{ u (u+1)}~,
 		\\
 		q(u)&=\frac{1}{3}\,\frac{39+4 u (7 u+5)}{3+ 4u(7 u+5)}~.
 	\end{split}
 	\ee

 	\subsection{Regions $I$ and $II$}
 
 	Starting with region $I$,  while focusing in  the case of large volume limit and small non-perturbative contributions, 
 	it can be observed that the requirement of a
 	positive second derivative of the potential at the minimum yields 
 	\be 
 	{\cal V}_{min}>\,\eta\,p(u)\Rightarrow \, -\eta\,p(u)W_0 \,>\,\eta\,p(u)~.
 	\ee 
 	From the range of $u\le -1$  (region $I$,  Table 1), it is deduced that 
 	$p(u)>0$ and taking into account the bound $ W_0\ge -1$  (for real values of the Lambert function), this implies that $\eta p(u)<0$ or $\eta <0$. Furthermore,  real $W_0$ values   defined  in~(\ref{VolVefmin}) imply that its argument
 	should be greater that $-e^{-1}$, which, for $\eta p(u)<0$ is 
 	satisfied for any $\xi,\eta$.
 	To determine whether a dS vacuum is attainable, the value of the effective potential at the minimum is required. 
 	A straightforward computation yields 
 	\be 
 	\begin{split}
 		\label{Vatmin}
 		V_{\rm eff}({\cal V}_{min} )&=\left(\tilde\epsilon {\cal W}_0\right)^2 
 		\frac{\eta (1+2 u) (3+14 ) -8 {\cal V}_{min} u (1+u)}{48\, {\cal V}_{min} ^3 }
 		\\
 		&=-\left(\tilde\epsilon {\cal W}_0\right)^2 \frac{u(1+u)}{6{\cal V}_{min}^3}\left({\cal V}_{min}-\frac{2}{3}\eta p(u) \right)~.
 	\end{split}
 	\ee
 	Taking into account  that for the range of $u\in(-\infty, -1)$ 
 	the  factor $u(1+u)>0$, it is readily seen that for the parameter space of region $I$ the value of the minimum (\ref{Vatmin}) is always negative. Hence when only F-term contributions are  taken into account, the resulting potential always exhibits an AdS vacuum. 
 	
 		\begin{figure}[H]
 		\centering
 		\includegraphics[scale=0.55]{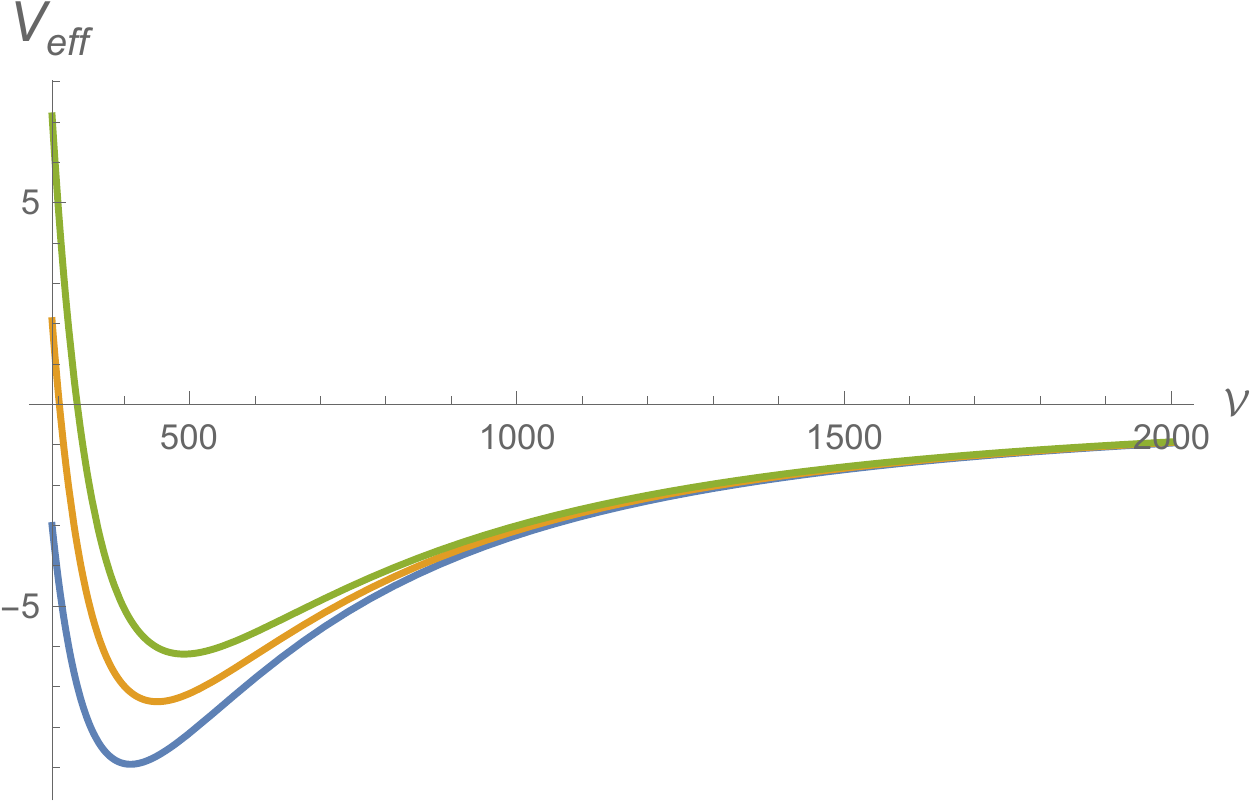}\;\;
 		\includegraphics[scale=0.5]{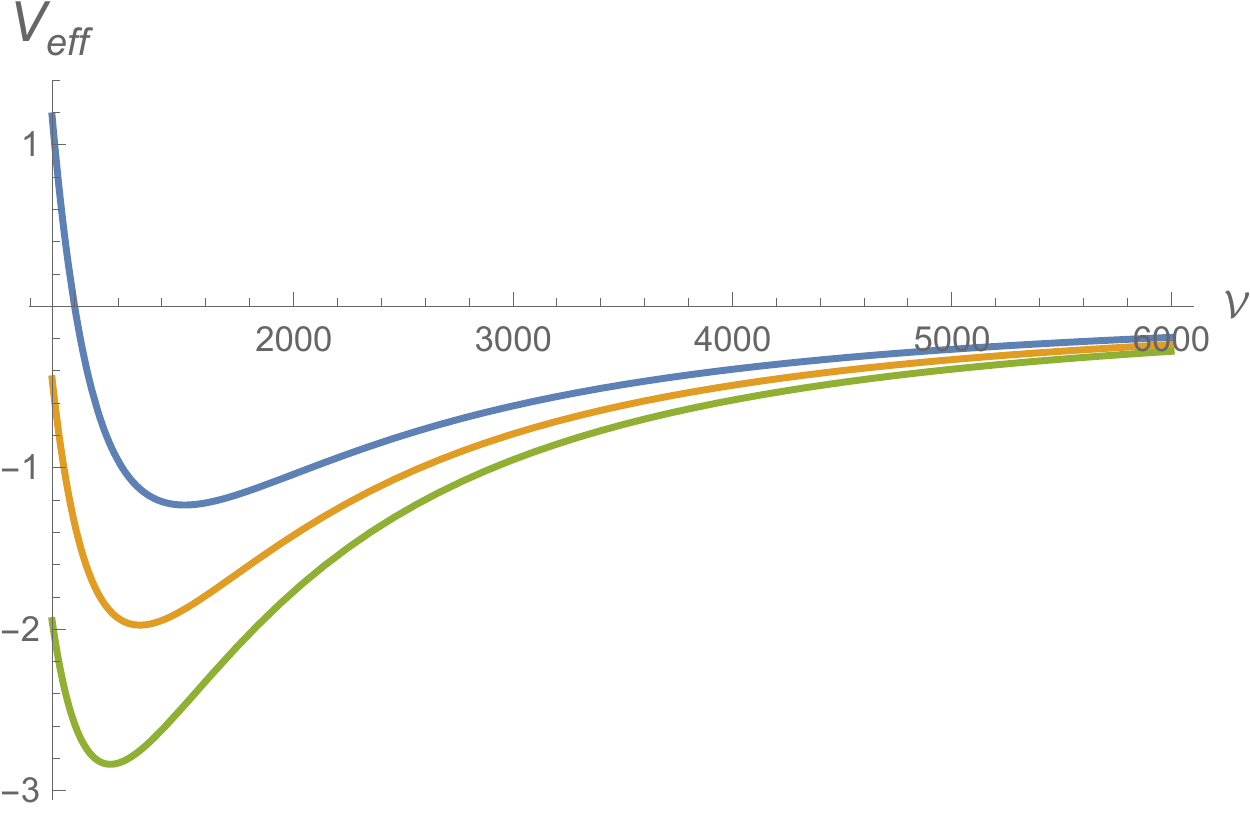}\;
 		\caption{Left panel: The F-term potential $V_{F}$ for  $\eta=-0.5,u=-9$ and 
 			three values of $\xi =150, 165, 180$. Lower $\xi$ values imply deeper AdS minima. Right panel: $V_{F}$ for  $\eta=-0.1, \xi= 200$ and three values of 
 			$u=-1.2,-1,25,-1.3$. The larger the $|u|$ values the deeper the AdS minima.
 		} 
 		\label{f3} 
 	\end{figure}
 	
 	The F-part of the potential is plotted in  figure~\ref{f3} 
 	for two values of the parameter $\eta$ and several values of $u=-b\tau_2$. 
    As expected, in all these cases the F-term potential implies always an AdS minimum and
 		an uplift term such as the one coming from a $\overline{D3}$-brane 
 		or D-terms induced form possible $U(1)$'s associated with   $D7$-branes is necessary.

 	\section{Uplifting with D-terms}

 	In the original KKLT construction~\cite{Kachru:2002gs} a dS vacuum was obtained  by including in the scalar potential an uplifting term 
 	$V_{up}\sim d/\tau^{3/2}$  whose  origin came from a $\overline{D3}$ brane.  It was pointed	out, however, that its presence breaks 
 	supersymmetry through  the $\overline{D3}$ decay due to the annihilation from fluxes  carrying $D3$ charge.  In the  framework of 
 	the four-dimensional effective supergravity theory an easy way of remedying these shortcomings  is to introduce a nilpotent 
 	supermultiplet $S$ \footnote{see for example~\cite{Ferrara:2014kva} and references therein}  associated 	with a Volkov-Akulov goldstino~\cite{Volkov:1973ix} (see for example~\cite{Rocek:1978nb,Komargodski:2009rz,Antoniadis:2010hs} for constraints on  $S$)
 	on the world volume of the 	$\overline{D3}$ brane.  Incorporating such a field in the present 
 	model, the superpotential and the K\"ahler potential	are modified as follows:
 	\begin{eqnarray}
 		\mathcal{W}&=&\mathcal{W}_0+Ae^{-a \tau_1}+Be^{-b \tau_2}+\mu^2 S~,\label{WithS}
 		\\
 		\mathcal{K}	&=&-2\log(\sqrt{\tau_1\tau_2\tau_3}+\xi+\eta \log(\tau_1\tau_2\tau_3)) +S\bar{S}~,\label{KWithS}
 	\end{eqnarray}
 	where  $\mu$ is an order one constant. However, with this modification, in the present model a
 	 dS minimum can be guaranteed only for the volume modulus. Along  the transverse  directions,  there are tachyonic fields 
 	 unless appropriate  uplifting terms from other sources are included. In the present  geometric setup this is naturally realised 
 	  by virtue of the  D-term contributions arising from the universal $U(1)$ factors of the $D7$ brane stacks.  
 	   The  uplifting of the AdS  to a dS vacuum takes place when  fluxes are turned on the associated $U(1)$ gauge fields
 	  and has been originally suggested
 	  by the authors of~\cite{Burgess:2003ic}.  The D-terms generated by the world-volume fluxes along the D7 brane stacks correspond
 	  to a shift symmetry   $\rho_j\to \rho_j +i Q_j \epsilon$  where $\rho_j$ are  the corresponding K\"ahler moduli and $Q_j$ the associated ``charges''. 
  In the presence	of  $\Phi_j$  fields which transform linearly under the corresponding abelian symmetry  with $q_j$ `charges',
 the flux induced D-terms	 acquire the general form~\cite{Haack:2006cy}
 	\be 
 	V_{D}= \frac{g_{D7_i}^{2}}{2}\left(Q_i\partial_{\rho_i} K+\sum_j q_j |\Phi_j|^2\right)^2,\; \frac{1}{g_{D7_i}^{2}}={\rm Im}\rho_i+\cdots~,  \label{Dterm}
 	\ee 
 	where   $\{\cdots\}$ stand for flux and dilaton dependent corrections.  For simplicity it will be assumed 
 	that the fields $\Phi_j$ have zero-vevs, hence, the generic form of the corresponding D-term potential becomes~\cite{Antoniadis:2018hqy}
 	\be 
 	{V}_{ \cal D}\; = \;\sum_{i=1}^3\frac{d_i}{\tau_i} \left(\frac{\partial { {\cal K}}}{\partial {\tau_i}}\right)^2
 	\;\approx \;
 	\sum_{i=1}^3 \frac{d_i}{\tau_i^3}\;\equiv\; \frac{d_1}{\tau_1^3}+\frac{d_2}{\tau_2^3}+\frac{d_3\tau_1^3\tau_2^3}{{\cal V}^6}~,\label{VDappr}
 	\ee 	
where the $d_i$ for $i=1,2,3$ are  positive constants  and are proportinal to   the  charges $Q_i^2$ as can be seen from~(\ref{Dterm}).
 	In the subsequent analysis the  case of large $\tau_1,\tau_2$ moduli will be considered (i.e. $a \tau_1\gg 1$ and $b \tau_2\gg 1$ )  where the calculations for the stabilization of the directions  transverse to the volume are simplified. This will provide  a more quantitative comparison of the effect of the ``strong" non-perturbative correction to the logarithmic one. Proceeding the way described above, the total potential is written as:
 	
 	\begin{equation}
 V_{\rm eff}\approx 	(\varepsilon W_0)^2 	\frac{7 (\xi +2 \eta  \log ({\cal V}))-4 {\cal V}}{8 {\cal V}^3}+
  \frac{d_1}{\tau_1^3}+\frac{d_2}{\tau_2^3}+\frac{d_3\tau_1^3\tau_2^3}{{\cal V}^6}~\cdot
  \label{VeffD}
 	\end{equation}
 	Minimization of~\eqref{VeffD} with  respect to the $\tau_1,\tau_2$ moduli, leads to the following equations:
 	\begin{align}
 		\tau_1^3&=\dfrac{d_1^{\frac{2}{3}}{\cal V}^2}{(d_2d_3)^{\frac{1}{3}}}~,\\
 		\tau_2^3&=\dfrac{d_2^{\frac{2}{3}}{\cal V}^2}{(d_1d_3)^{\frac{1}{3}}}~\cdot
 	\end{align}

\begin{figure}[H]
	\centering\includegraphics[scale=0.45]{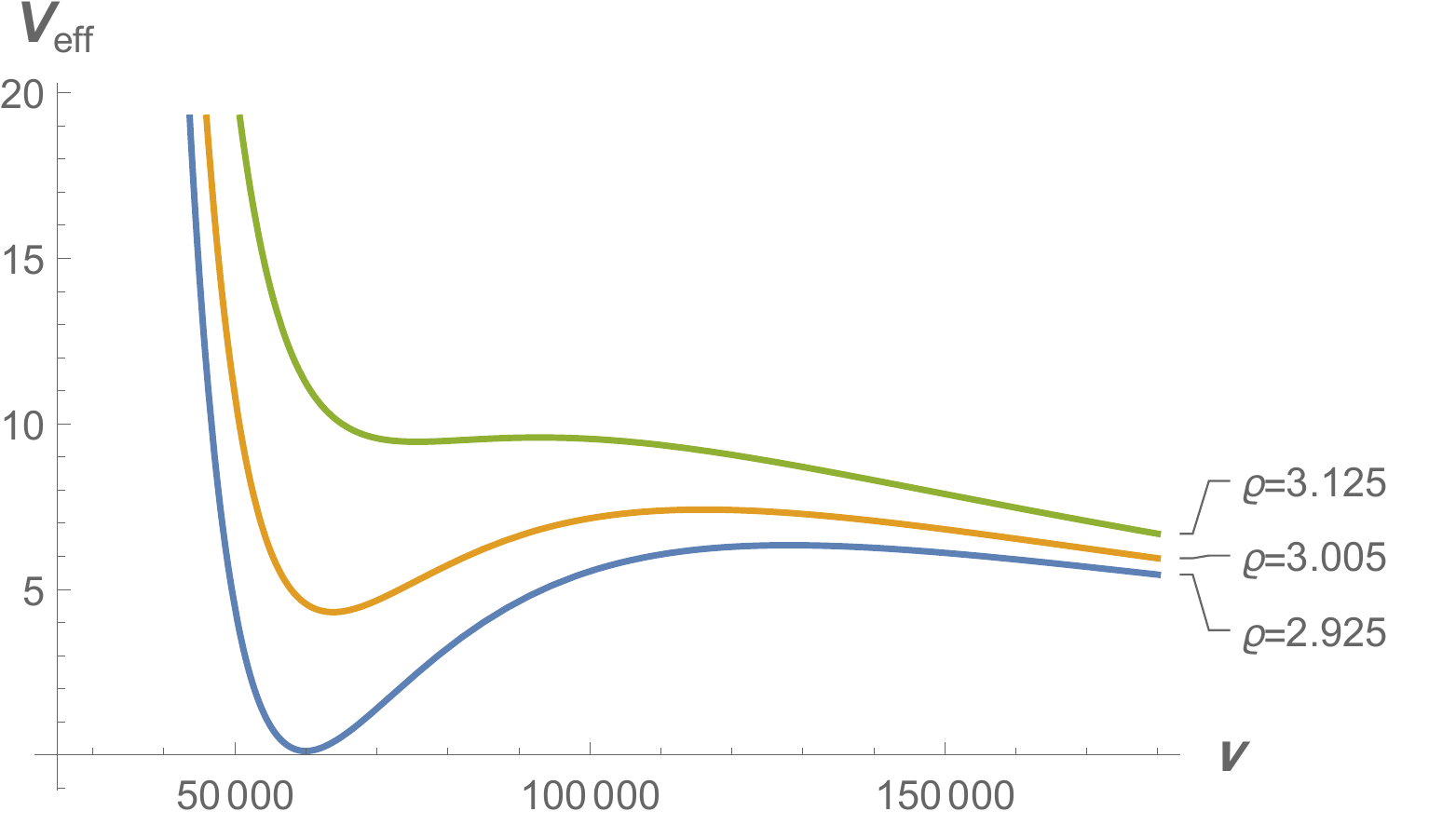},
	\caption{The potential (\ref{VwithD}) for $\eta=-0.5,\xi=10$  and three values of the parameter $\varrho =10^{5}(6 r-1)$.  For $\varrho =2.925$ the potential at the minimum vanishes. For larger $\varrho$ values $V_{\rm eff}({\cal V}_{min})>0$ while the minimum disappears for $\varrho \gtrsim 3.125$. }
	\label{rhominmax}
\end{figure}

\begin{figure}[H]
\centering\includegraphics[scale=0.35]{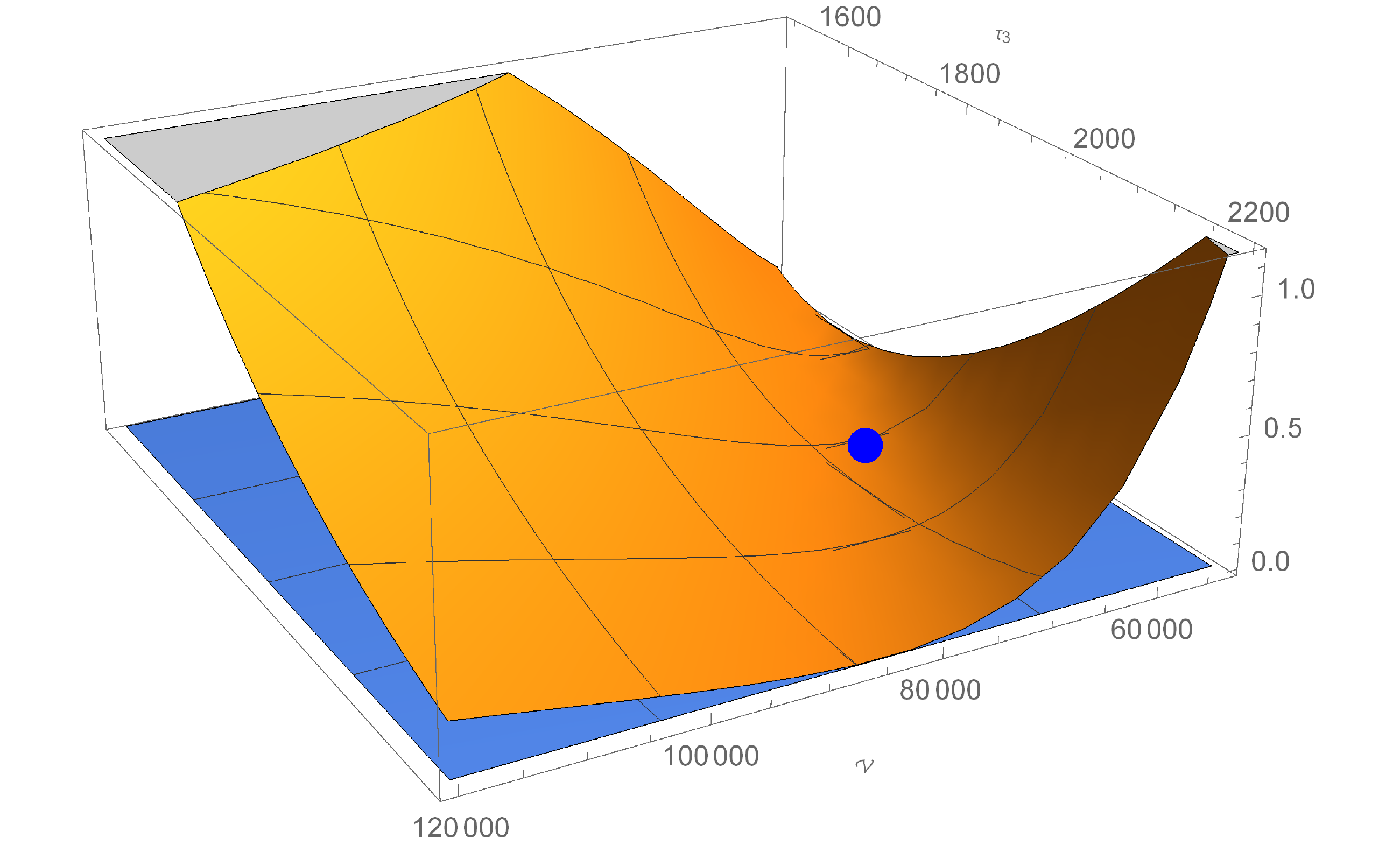}
 \caption{The potential (\ref{VwithD}) for $\eta=-0.5,\xi=10,d=0.6668,\mathcal{W}_0=-1,\varepsilon\approx 2,\varrho=3.05,d_2=d_3=1$. The  light blue plane is just above $V=0^+$ and  the blue dot is the intersection with $V_{\rm eff}$ which indicates the position of the dS the minimum.}
 \label{3dpot}
\end{figure}

Substituting the above back into~(\ref{VeffD}), the potential $V_{\rm eff}$ receives the following compact formula:

\begin{align}
	V_{\rm eff} \approx({\eps}\mathcal{W}_0)^2
	\left(\frac 78\dfrac{\xi+2\eta \log(\mathcal{V})}{\mathcal{V}^3}-\frac{1}{2{\cal V}^2}\right)+\frac{3d}{{\cal V}^2}~,
	\label{VwithD}
\end{align}
where $d=(d_1d_2d_3)^{1/3}$.
The volume modulus at the minimum of the potential is
\begin{align}
\mathcal{V}_{min}=\dfrac{21\eta}{4(6 \mathtt{r}-1)}W_0\big(\dfrac{4(6\mathtt{r}-1)}{21\eta}e^{\frac{1}{3}-\frac{\xi}{2\eta}}\big),\label{minvol}
\end{align}
where the new parameter $\mathtt{r}$ introduced in the formula of  $\mathcal{V}_{min}$ above is the ratio of the
F- and D-term coefficients
\be 
 \mathtt{r}=\dfrac{d}{(\varepsilon \mathcal{W}_0)^2}~.\label{DFratio}
 \ee
For given $\xi$ and $\eta$ the   coefficient $\mathtt{r}$   has an upper and 
a lower bound  coming from the following two constraints:  
i) Real values of the volume  are achieved when  the argument of the 
$W_0$
function must be larger than $-1/e$ and ii) the potential at the minimum must be positive. Implementing these conditions, the following bounds
on $\mathtt{r}$  are imposed
\begin{align}
\dfrac{1}{6}+\frac{7}{12}\frac{|\eta|}{{\cal V}}&\lesssim\mathtt{r}\lesssim\dfrac{1}{6}+\dfrac{7|\eta|}{8}e^{-\frac{\xi}{2|\eta|}-\frac{4}{3}}~.
\label{r2bounds}
\end{align}	
For positive and large $\xi$ values, this restricts the values of $\mathtt{r}$ in a tiny region close to $\frac 16$.  It should be  observed that the exact value $\mathtt{r}=\frac 16$ eliminates the $\frac{1}{{\cal V}^2}$ term form the scalar potential. This would leave only the perturbative F-part $\propto (\xi+2\eta \log{\cal V})/{\cal V}^3$  which defines only AdS minima.  It is worth noticing that,  this value is twice as big compared with that obtained in the case of the effective potential~(\ref{Veff0})  derived with only one non-perturbative term  in the superpotential. It is
convenient to define a new parameter 
\be
  \varrho = 10^5 (6 \mathtt{r}-1)~,\label{rhoratio}
  \ee 
which can be used to plot the effective potential~(\ref{VeffD}).  Assuming for example the values $\xi=10, \eta=-0.5$ and using (\ref{r2bounds}), it can be deduced that a dS minimum exists as long as 
\[  2.925 \lesssim    \varrho \lesssim 3.125~. \]
The potential (\ref{VwithD}) is plotted in figure~\ref{rhominmax}  as a function of the volume for three values of 
the parameter $\varrho $. In figure~\ref{3dpot} a three dimensional plot   is shown where  the  minimum is depicted along ${\cal V}$ and $\tau_1$ directions.

\section{Probing the strong coupling regime}

The approximations used in the previous sections were suitable to explore  regions of the parameter space where the moduli $\tau_{1,2}$ are large and the 
exponential factors  $e^{-a\tau_1}, e^{-b\tau_2}$ of the non-perturbative terms in the superpotential are suppressed. 
The analysis in this section will also be valid   for regions of moduli values not much larger than  $a\tau_1\sim  b\tau_2\gtrsim  {\cal O}(1)$   where the non-perturbative corrections are substantially larger.
To this end, a different approach will be followed in this section and the solutions of
the flatness conditions 
(\ref{flatW1},\ref{flatW20}) will be expressed in terms of 
 the parameter $\gamma$  and the modulus $\tau_2$, namely:
\begin{equation}
	\begin{split}
		b\tau_2&=-W_{0/-1}\big(-\dfrac{a \tau_1 e^{a \tau_1}}{\beta}\big)\equiv {\mathbf{w}}(\tau_1)~,
		\\
			\gamma&=-(1+2a\tau_1) e^{-a\tau_1}-\beta e^{-b \tau_2}~,
		\\
		&=-(1+2a\tau_1) e^{-a\tau_1}-\beta e^{ -{\mathbf{w}}(\tau_1)}~,
	\end{split}
\label{2flatc}
\end{equation}
where ${\mathbf{w}}(\tau_1)$ is shorthand for the  $b\tau_2$ solution.
It should be remarked that, in contrast to the first case,  these solutions of (\ref{flateq2}) are exact  and no assumptions have been made. The only drawback of this parametrization is that the solutions of the minimization conditions are expressed in terms of implicit functions for all moduli fields
and thus, it is more difficult to handle them analytically.
 Imposing the two conditions (\ref{2flatc}), the scalar potential takes the simple form 
\begin{equation}
	V_{F}= (4 a)^2 A^2 \tau_1^2 e^{-2 a \tau_1}\left(\frac 78 \frac{ \xi+ 2 \eta \log (\mathcal{V}) }{\mathcal{V}^3}
	-\frac{1}{2\mathcal{V}^2}\right)
	-\eta \xi\frac{68 a^2 A^2   \tau_1^2 e^{-2a \tau_1 } \log (\mathcal{V})}{\mathcal{V}^4}~.
	\label{Veff4} 
\end{equation}

\begin{figure}[H]
	\centering\includegraphics[scale=0.65]{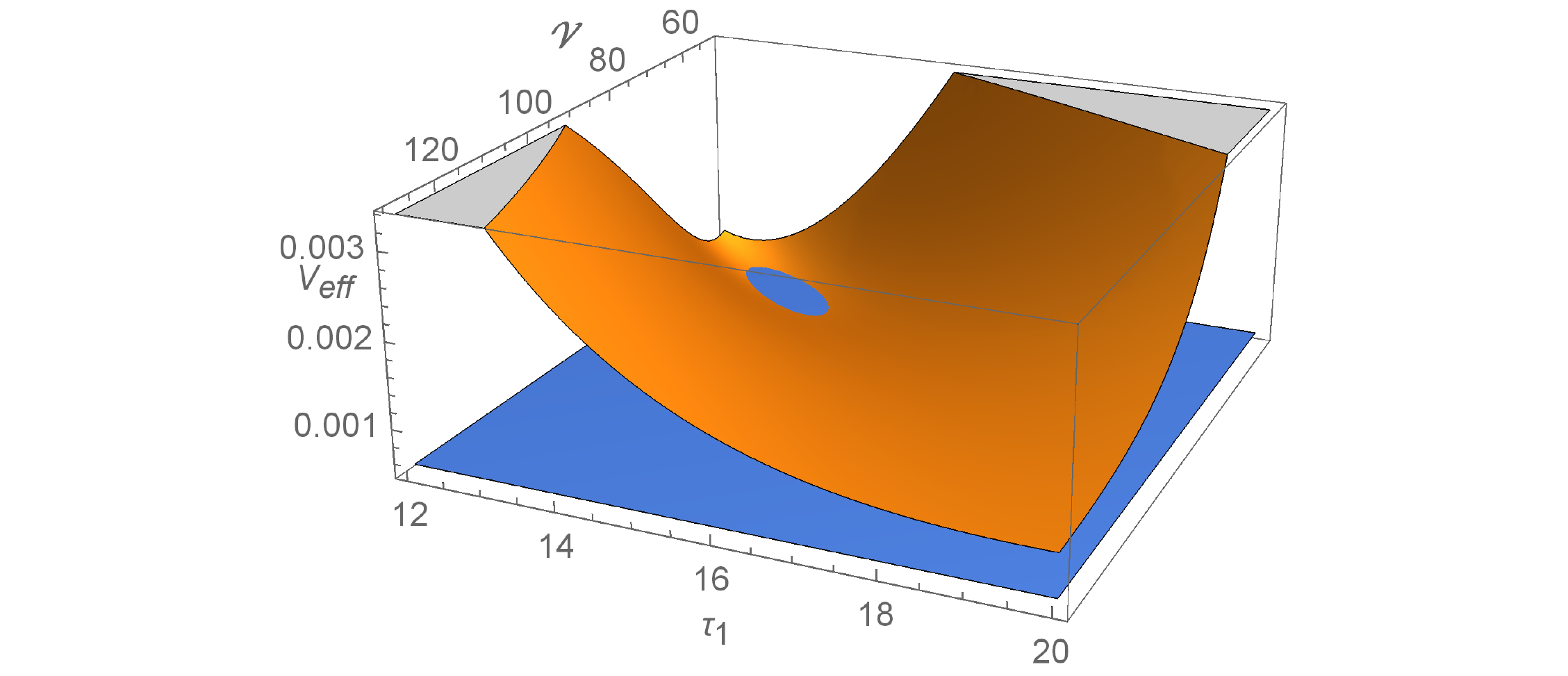}
	\caption{The minimum of the potential (\ref{VFDex1}) in regions with  moderate values $\tau_i$. Again, as in the previous figures, the blue spot indicates the position of the dS minimum.}
	\label{LargeNP}
\end{figure}

\noindent This potential encapsulates all the features of the potentials  derived in the previous sections: the pure non-perturbative term $\sim 1/\mathcal{V}^2$, the perturbative one $\sim  1/\mathcal{V}^3$ and a higher order mixing term $\sim 1/\mathcal{V}^4$. Minimization along the volume direction gives the value of ${\cal V}$ at the minimum of the potential:
\begin{equation}
	\mathcal{V}_{min}\approx -\frac{21\eta}{4}  W_0\left(-\frac{4 e^{\frac{1}{3}-\frac{\xi }{2 \eta}}}{21 \eta}\right)~.
\end{equation}
It was already  pointed out in the previous sections that the F-term part implies always an AdS minimum  
 and thus the  D-terms are crucial for its uplifting to a dS vacuum. Including the D-term part~(\ref{VDappr}), the total potential is written as:
\be
	\label{VFDex} 
\begin{split}
	V_{\rm eff}&=V_F+V_D\\
	&\approx(4 a)^2 A^2 \tau_1^2 e^{-2 a \tau_1}\left(\frac 78 \frac{ \xi+ 2 \eta \log (\mathcal{V}) }{\mathcal{V}^3}
	-\frac{1}{2\mathcal{V}^2}\right)+\dfrac{d_1}{\tau_1^3}+\dfrac{d_3}{\tau_3^3}+\dfrac{d_2 \tau_1^3\tau_3^3}{\mathcal{V}^6}~,
\end{split}
\ee
where in the second line the $\frac{1}{{\cal V}^4}$ from $V_F$ has been omitted.  Minimizing (\ref{VFDex}) with respect to the $\tau_3$ modulus, its minimal value along $\tau_3$ is found  to be:
\begin{align}
	\tau_3=\left(\dfrac{d_3}{d_2}\right)^{1/6}&\dfrac{\mathcal{V}}{\sqrt{\tau_1}}~.\label{t3sol}
\end{align}
Substitution of~(\ref{t3sol}) into~(\ref{VFDex}) yields,
\be 
	V_{\rm eff}=(4 a)^2 A^2 \tau_1^2 e^{-2 a \tau_1}\left(\frac 78 \frac{ \xi+ 2 \eta \log (\mathcal{V}) }{\mathcal{V}^3}
	-\frac{1}{2\mathcal{V}^2}\right)+\dfrac{d_1}{\tau_1^3}+\dfrac{2\sqrt{d_2d_3}\tau_1^{3/2}}{\mathcal{V}^3}~.\label{VFDex1}
 \ee 
 Once a positive minimum along the volume direction is ensured, with a few more additional constraints on the parameter space, 
 it can be shown that a  minimum along all moduli directions is achievable. 
In figure~\ref{LargeNP} a characteristic case is plotted where 
the moduli fields acquire moderate values.

\section{Concluding remarks}

The cosmological predictions of effective  theory models derived from string theory compactifications are the subject of intensive investigation.
Among the primary and most important objectives  towards this goal is the construction of  string vacua  with  positive cosmological
constant.  In this work  the combined effects of novel perturbative logarithmic
string loop corrections in the K\"ahler potential, and non-perturbative contributions in the superpotential  have been analyzed in the framework
of type IIB  string theory.  In general, such quantum corrections   are of key significance  not merely for  the existence 
 of de Sitter (dS)  minima but also  for  the (well known)  moduli stabilization problem. 
 Type IIB  string theory and  in particular its geometric variant,   F-theory,   are  of  great interest  since they provide the necessary ingredients for a successful  interpretation  of  these two important  issues.  Within this framework, a geometric 
 configuration of three $D7$ brane stacks intersecting each other is considered  and 
 three K\"ahler moduli are introduced whose imaginary parts $\tau_i$  determine 
 the internal volume  through ${\cal V}=\sqrt{\tau_1\tau_2\tau_3}$.
 
 The  logarithmic terms involve the K\"ahler moduli and are induced from loop corrections when  closed strings emitted from localized Einstein-Hilbert (EH) terms
 propagate through the codimension-two volume towards the seven-brane probes~\cite{Antoniadis:2019rkh}.
 Such novel EH terms originate from the $R^4$ corrections of the 
 effective ten-dimensional string  action and appear only in four spacetime dimensions. The non-perturbative effects are assumed to be associated with gaugino condensation on the $D7$ stacks and modify the superpotential 
 with corrections of the usual exponential K\"ahler moduli dependence. 
  The case of non-perturbative corrections where one or two K\"ahler 
 moduli contribute has been worked out. In the large volume regime it has been found that  the potential   takes the generic simple form
 \[  V_{\rm eff} = a\frac{\xi + \eta\log{\cal V}}{{\cal V}^3}+\frac{b-c}{{\cal V}^2}~,
 \]
 where $a,b,c$ positive coefficients and the parameters $\xi$ -proportional to the Euler characteristic- and $\eta=-\xi g_s T_0/2$ with $T_0$ the $ D7$ brane tension and $g_s$ the string coupling~\cite{Antoniadis:2019rkh}. The term  proportional to $1/{\cal V}^3$ encapsulates  perturbative  contributions through the constants $\xi,\eta$ and the term $(b-c)/{\cal V}^2$   combines positive contributions from D-terms and negative ones from non-perturbative corrections. The necessary conditions for  K\"ahler moduli  stabilization and dS minima are  $\eta<0$ ($\xi>0$) and  $b>c$. It has been demonstrated in this article that a simultaneous  solution for both problems can be guaranteed for a variety of appropriate flux choices and gaugino condensation parameters.  
 
Finally, it should be noted  that within the  present geometric setup, there may exist  charged matter  fields associated with the excitations of open strings with their ends  attached to $D7$  brane stacks. Their scalar components  may receive supersymmetric positive square masses  from brane separation or Wilson lines,  and non-supersymmetric contributions due to the presence of the worldvolume magnetic fields. 
With suitable conditions on  various quantities such as  magnetic fluxes and geometric characteristics (e.g. ratios of
torii areas),  all but a few  masses-squared turn positive. In the simplest scenario it can be arranged that one tachyonic field  arises from magnetised D7 brane  identified with its orientifold image. In fact, its mass squared  turns negative when the internal volume acquires a critical value.  
As a consequence, they induce specific contributions to the F- and D-terms of the effective potential.   With these new ingredients,  
it has been shown~\cite{Antoniadis:2021lhi} that this can play the r\^ole of the waterfall field, 
 providing in this way an explicit string realization of the hybrid inflationary scenario. 
 
 \vspace*{1cm}
 {\bf Acknowledgments:} 
 {\it ``This research work was supported by the Hellenic Foundation for Research and Innovation (H.F.R.I.) under the ``First Call for
 	H.F.R.I. Research Projects to support Faculty members and	Researchers and the procurement of high-cost research equipment
 	grant'' (Project Number: 2251)''}.  {\it The work of V.B. is partially supported 	by the Research Committee of the University of Ioannina.}

\newpage

 \section{ Appendix}

 \subsection{ The  Potential with a single non-perturbative correction}
 The effective potential with a single non-perturbative correction $Ae^{i \alpha \rho_1}$  has the following form:
 \begin{equation}
 	V_{\rm eff}=(\epsilon \mathcal{W}_0)^2\dfrac{2\xi-\mathcal{V}+4\eta((\log(\mathcal{V})-1)}{4\mathcal{V}^3}+\dfrac{d_1}{\tau_1^3}+\dfrac{d_2}{\tau_2^3}+\dfrac{d_3}{\tau_3^3},\;\; \epsilon=\dfrac{4}{2+\frac{1}{\alpha \tau_1}}~.
 	\label{V1NP} 
 \end{equation}
 Exchanging one modulus e.g. $\tau_2$ with the volume ${\cal V}$   and minimizing for $\tau_3$, (\ref{V1NP}) takes the form:
 \begin{equation}
 	V_{\rm eff}=(\epsilon \mathcal{W}_0)^2\dfrac{2\xi-\mathcal{V}+4\eta(\log(\mathcal{V}-1)}{\mathcal{V}^3}+\dfrac{d_1}{\tau_1^3}+\dfrac{2d\tau_1^{3/2}}{\mathcal{V}^3},\;\; d=\sqrt{d_2d_3}~,
 \end{equation}
where 
 \begin{equation}
	\tau_3=(\dfrac{d_3}{d_2})^{1/6}\dfrac{\mathcal{V}}{\sqrt{\tau_1}}~.
\end{equation}
Next, minimization with respect to the $\tau_1$,  leads to the following equation:
 \begin{align}
 	(3\tau_1^{9/2}d-3d_1\mathcal{V}^3)(1+2\alpha\tau_1)^3-8(\alpha\mathcal{W}_0)^2\tau_1^5(\mathcal{V}-2\xi+4\eta(1-\log(\mathcal{V}))=0~.\label{eq}
 \end{align}
 \noindent
For large $\mathcal{V}, \tau_1$ and in the limit where the second term can be ignored,  this equation takes the simple form 	$(3\tau_1^{9/2}d-3d_1\mathcal{V}^3)\approx 0$ 
and the solution is 
 \begin{align} \tau_1\approx (\dfrac{d_1^2}{d_2d_3})^{1/9}\mathcal{V}^{2/3}~.\label{sol}
 \end{align}
The potential at the minimum is
 \begin{equation}
 	V_{\rm eff}=\dfrac{3\tilde{d}}{\mathcal{V}^2}+\dfrac{4\mathcal{W}_0^2(-\mathcal{V}+2\xi+4\eta(\log(\mathcal{V})-1)}{g^2\mathcal{V}^3}~.\label{Veff1A}
 \end{equation}
Thus, it takes the previous form up to the factor  $g=2+\dfrac{\tilde{d}^{1/3}d_1^{-1/3}}{\alpha \mathcal{V}^{2/3}}$, whereas
 $\tilde{d}=(d_1d_2d_3)^{1/3}$.
 For large ${\cal V}$,  $g\approx 2$ and $\log(\mathcal{V})\gg 1$ the potential (\ref{Veff1A})  reduces to that of (\ref{Veff1}).

 \subsection{The F-term potential with two NP corrections}

The exact form of the perturbative part of the F-term potential in section 3 is
 \be 
 V_F^{(p)}=
 \frac{3 {\cal W}_0^2 (-4 \eta  (\eta +\xi )+	{\cal V} (\xi -8 \eta )+2 \eta  (	{\cal V}-4 \eta ) 
 	\log (	{\cal V}))}{(\xi +2 \eta  \log (	{\cal V})+	{\cal V})^2 \left(4 \eta  (3 \eta +\xi )+2
 	{\cal V}^2+	{\cal V} (16 \eta -\xi )-2 \eta  (	{\cal V}-4 \eta ) \log (	{\cal V})\right)}~.\nonumber 
 \ee 
 Expanding in $\eta$ and $\xi/{\cal V}$, this takes the simplified form
 \be 
 \label{FapproxA}
 \begin{split}
 	V_F^{(p)}&\approx 
 	\frac{3 {\cal W}_0^2 ({\cal V} (\xi -8 \eta )+2 \eta  ({\cal V}-3 \xi ) \log ({\cal V}))}{2 {\cal V}^4}
 	\\
 	&\approx 
 	\frac{3}{2} {\cal W}_0^2 \frac{\xi +2 \eta  \log ({\cal V})}{ {\cal V}^3}+{\cal O}(\frac{1}{ {\cal V}^4})~.
 \end{split}
 \ee 
 The last approximated form~(\ref{FapproxA})  is the expression~(\ref{Fpapprox})  in section 3.

 \noindent The pure non-perturbative part $	V_F^{(np)}$ is 
 
 \[\frac{{a \tau_1}  A  e^{-{a \tau_1}}\left( (1+{a \tau_1})  A e^{- {a \tau_1}}+ \left({\cal W}_0+B e^{-b \tau_2}\right)\right)+b\tau_2 B  e^{-b \tau_2}\left( \left({\cal W}_0+A e^{-{a \tau_1}}\right)+ B 
 	e^{- b \tau_2} (1+b \tau_2)\right)}{{\cal V}^2/4}~\cdot \]
 Implementing the definitions ${\cal W}_0/A= \gamma, B/A= \beta $, and 	making successively the substitutions  
 \be  
 \gamma \to 2 we^{-a\tau_1},\, a\to -(1 + 2 w)/(2  \tau_1),\, b\to -u/\tau_2~,\label{wtaurelationsA}
 \ee 
 which follow from the solutions~(\ref{Lambert}),  this 
 becomes
 \[	V_F^{(np)}=   -{\cal W}_0^2 \frac{(u+1) (2 w+1)^2}{2 u w^2 {\cal V}^2}~,\]
 which is the form~(\ref{FNPpure}) of section 3. 
 
 Remarkably, this term has a volume dependence $\propto  \frac{1}{{\cal V}^2}$ which is exactly the dependence  of the D-term uplift in~\eqref{Veff1}.
 For the regions $I,II$ where the approximation is valid,	 however, because $u (1+u)>0$ the contribution of this term 
 is negative and deepens the AdS vacuum.\footnote{ Nonetheless, it will be seen that this term  has the same power-law volume   dependence with the positive D-term contributions $d/{\cal V}^2$ and can be compensated by larger values of the  parameter $d$.}
 
 The mixing term of perturbative and non-perturbative corrections is a complicated
 function and 	can be elaborated in a similar manner.

\subsection{The exact potential}
In order to probe various regions of the parameter space, in the main text we have worked out several approximations of the F-term  scalar potential. These constitute limiting cases of  exact form  given below:
 
 $$V_F^{ex.}=\frac{e^{-2 \left(a \tau _1+b \tau _2\right)} \left(2 B N_3 e^{a \tau _1+b \tau _2}+B^2
 	N_1 e^{2 a \tau _1}+N_2 e^{2 b \tau _2}+2 \eta  N_4 \log ( {\cal V})\right)}{(2 \eta + {\cal V}) (\xi +2 \eta  \log ( {\cal V})+ {\cal V})^2 \left(12 \eta ^2+4 \eta  (\xi +4  {\cal V})+2 \eta  (4
 	\eta - {\cal V}) \log ( {\cal V})+ {\cal V} (2  {\cal V}-\xi )\right)}~,$$
where 
 \be 
 \begin{split}
 	N_1&=
 2 \eta  \left(8 b^2 \xi ^2 \tau _2^2+4  {\cal V}^2 \left(2 b \tau _2 \left(3 b \tau
 _2+2\right)-3\right)+\xi   {\cal V} \left(16 b \tau _2 \left(2 b \tau
 _2+1\right)-3\right)\right)-24 \eta ^3
 \\&+ {\cal V}^2 \left(8 b \tau _2 \left(b \tau _2+1\right) (\xi
 + {\cal V})+3 \xi \right)+4 \eta ^2 \left(8 b \xi  \tau _2 \left(b \tau _2+1\right)+ {\cal V}
 \left(8 b \tau _2 \left(b \tau _2+1\right)-15\right)-6 \xi \right)~,
 \end{split}
\nn
\ee
 \be 
\begin{split}
 	N_2&=A^2 \left(2 \eta  \left(8 a^2 \xi ^2 \tau _1^2+4  {\cal V}^2 \left(2 a \tau _1 \left(3 a \tau
 	_1+2\right)-3\right)+\xi   {\cal V} \left(16 a \tau _1 \left(2 a \tau
 	_1+1\right)-3\right)\right)\right.
 	\\&\left. + {\cal V}^2 \left(8 a \tau _1 \left(a \tau _1+1\right) (\xi
 	+ {\cal V})+3 \xi \right)+4 \eta ^2 \left(8 a \xi  \tau _1 \left(a \tau _1+1\right)+ {\cal V}
 	\left(8 a \tau _1 \left(a \tau _1+1\right)-15\right)-6 \xi \right)-24 \eta
 	^3\right)
 	\\&-2 A {\cal W}_0 e^{a \tau _1} (2 \eta + {\cal V}) \left(4 \eta  \left(-2 a \xi  \tau _1-2 a
 	\tau _1  {\cal V}+3 \xi +6  {\cal V}\right)-{\cal V}\left(4 a \tau _1 (\xi + {\cal V})+3 \xi \right)+12 \eta
 	^2\right)
 	\\&-3 {\cal W}_0^2 e^{2 a \tau _1} (2 \eta + {\cal V}) \left(4 \eta ^2+4 \eta  (\xi +2  {\cal V})-\xi 
 	 {\cal V}\right)~,
 	 \end{split}
 \nn
\ee
\be 
\begin{split}
 	N_3&=A \left(-4 \eta ^2 \left(2 \xi  \left(2 b \tau _2 \left(a \tau _1-1\right)-2 a \tau
 	_1+3\right)+{\cal V} \left(4 b \tau _2 \left(a \tau _1-1\right)-4 a \tau
 	_1+15\right)\right)\right.\\&\left.
 	+2 \eta  {\cal V} \left(\xi  \left(-8 a b \tau _2 \tau _1+8 a \tau _1+8
 	b \tau _2-3\right)+{\cal V} \left(-8 a b \tau _2 \tau _1+8 a \tau _1+8 b \tau
 	_2-12\right)\right)\right.
 	\\&\left.
 	+{\cal V} \left(4 a \tau _1 (\xi +{\cal V}) \left(b \xi  \tau _2+{\cal V}\right)+{\cal V}
 	\left(4 b \tau _2 (\xi +{\cal V})+3 \xi \right)\right)-24 \eta ^3\right)
 	\\
 	&-{\cal W}_0 e^{a \tau _1} (2
 	\eta +{\cal V}) \left(4 \eta  \left(-2 b \xi  \tau _2-2 b \tau _2 {\cal V}+3 \xi +6 {\cal V}\right)-{\cal V}
 	\left(4 b \tau _2 (\xi +{\cal V})+3 \xi \right)+12 \eta ^2\right)~,
 	 \end{split}
 \nn
\ee
\be 
\begin{split}
 		N_4&=e^{2 b \tau _2} \left(A^2 \left(32 a^2 \eta^2 \tau_1^2 \log ({\cal V})+\eta \left(32 a^2 \xi  \tau_1^2+{\cal V} (32 a \tau _1 (2
 		a \tau _1+1)-6)\right)+8 \eta^2 (2 a \tau_1-1) (2 a \tau_1+3)\right.\right.
 		\\&\left.\left. +{\cal V}^2 (8 a \tau _1 (a \tau _1+1)+3)\right)+2 A
 		{\cal W}_0 e^{a\tau _1} (2 \eta+{\cal V}) (4 \eta (2 a \tau _1-3)+{\cal V} (4 a \tau _1+3))
 		\right.
 		\\&\left.
 		-3 {\cal W}_0^2 e^{2 a \tau _1} (4
 		\eta-{\cal V}) (2 \eta+{\cal V})\right)+2 B e^{a \tau _1+b \tau _2} \left(A \left(-8 \eta^2 (2 b \tau _2 (a \tau _1-1)-2 a \tau _1+3)
 		\right.\right.
 		\\&\left.\left. 
 		+2 \eta
 		{\cal V}(-8 a b \tau _1 \tau _2+8 a t+8 b \tau _2-3)+8 a b \eta \tau _1 \tau _2 {\cal V} \log ({\cal V})+{\cal V} (8 a b \xi  \tau _1 \tau _2+{\cal V} (4 b \tau _2(a
 		\tau _1+1)+4 a \tau _1+3))\right)
 		\right.
 		\\&\left.
 	+	{\cal W}_0 e^{a \tau_1} (2 \eta+{\cal V}) (4 \eta (2 b \tau _2-3)+{\cal V} (4 b
 		\tau _2+3))\right)+B^2 e^{2 a \tau _1} \left(32 b^2 \eta^2 \tau _1^2 \log ({\cal V})
 		\right.
 		\\&\left.
 		+\eta \left(32 b^2 \xi  \tau _2^2+{\cal V}
 		(32 b \tau _2 (2 b \tau _2+1)-6)\right)+8 \eta^2 (2 b \tau _2-1) (2 b \tau _2+3)+{\cal V}^2 (8 b \tau _2 (b \tau _2+1)+3)\right)~.
 \end{split}
\nn
 \ee

\begin{figure}[H]
 	\centering\includegraphics[scale=0.45]{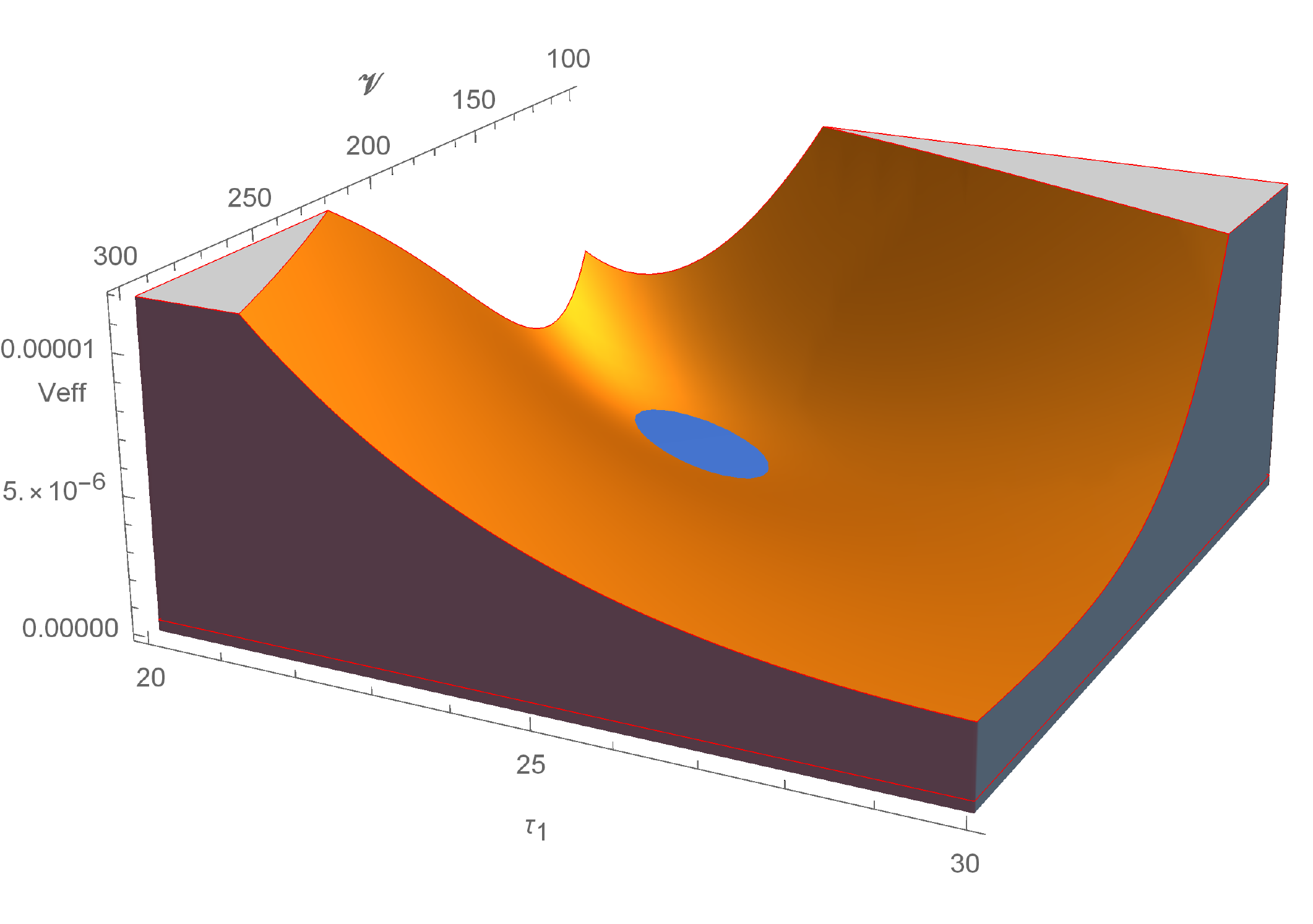}
 	\caption{Exact Potential (in arbitrary units) for the numerical coefficients~(\ref{valex}).   The blue spot is at a height  $V_0\approx  4\times 10^{-7}>0$ and defines the
 		position of the dS minimum.  }
 	\label{Vexactplot} 
 \end{figure}

 Using the exact potential, all regions of the parameter space   can be probed. For example,
 cases with $\xi$ comparable with ${\cal V}_{min}$ cannot be examined
 in the approximations discussed so far. Here is a case with large $\xi$. 
 The following values 	of the parameters 
 \be 
 \xi = 100,\, a= 0.1,\, b= 0.1,\, A= 1,\, B = 1,\, {\cal W}_0= -1.2,\,\eta= -1,\, d= 0.35,\, c=
 0.205~,\label{valex}
 \ee 
 are used to plot figure~\ref{Vexactplot} with the exact $V_{\rm eff}$.
 
\end{document}